\documentclass[a4paper,fleqn,usenatbib]{mnras}
\pdfoutput=1
\usepackage{graphicx}
\usepackage{array,amssymb,amsmath}
\usepackage[usenames]{color}
\usepackage{mathrsfs}
\usepackage{tikz}
\usepackage{times}
\usepackage{subfigure}

\usepackage{psfrag}
\newcommand{\nc}{\newcommand}

\newcommand{\be}{\begin{equation}}
\newcommand{\ee}{\end{equation}}
\newcommand{\bea}{\begin{eqnarray}}
\newcommand{\eea}{\end{eqnarray}}

\newcommand\Mpc{\,\mbox{Mpc}}

\newcommand\lsim{\mathrel{\rlap{\lower4pt\hbox{\hskip1pt$\sim$}}
    \raise1pt\hbox{$<$}}}
\newcommand\gsim{\mathrel{\rlap{\lower4pt\hbox{\hskip1pt$\sim$}}
    \raise1pt\hbox{$>$}}}
\def\ee{\end{equation}}
\def\be{\begin{equation}}
\def\ea{\end{align}}
\def\ba{\begin{align}}
\newcommand{\Omk}{\Omega_\kappa}

\newcommand{\Om}{\Omega_m}
\newcommand{\OL}{\Omega_\Lambda}

\newcommand{\dr}{\text{d}}

\newcommand{\Ode}{\Omega_\text{de}}

\newcommand{\Cp}{{\mathscr{C}}}

\newcommand{\mb}{m_B^*}
\newcommand{\mbi}{m_{Bi}^*}

\newcommand{\norm}{\mathcal{N}}

\newcommand{\normal}[3]{\mathcal{N}_{#1}\!\!\left( #2, #3 \right)}

\newcommand{\sx}{\sigma_x}

\newcommand{\sxi}{\sigma_{x,i}}
\newcommand{\syi}{\sigma_{y,i}}
\newcommand{\syxi}{\sigma_{yx,i}}

\newcommand{\xh}{\hat{x}}
\newcommand{\yh}{\hat{y}}

\newcommand{\xhi}{\hat{x}_i}
\newcommand{\yhi}{\hat{y}_i}
\newcommand{\whi}{\hat{w}_i}

\newcommand{\Rx}{R_x}
\newcommand{\Rc}{R_c}

\newcommand{\cbl}{\underline{c}}

\newcommand{\xbl}{\underline{x}_1}

\newcommand{\salt}[1]{{\sc SALT-II}}
\newcommand{\saltorig}[1]{{\sc SALT}}
\newcommand{\sifto}[1]{{\sc SiFTO}}
\newcommand{\mlcs}[1]{{\sc MLCS}}

\nc{\Cm}{\hat{C}}
\nc{\sigint}{\sigma_{\mu}^\text{int}}
\nc{\sint}{\sigma_{\text{int}}}
\nc{\sinti}{\sigma_{\text{int},i}}
\nc{\muth}{\mu^{\text{th}}}
\nc{\cnot}{c_\star}
\nc{\xnot}{x_\star}
\nc{\onesn}{\mathbf{1}_n}
\nc{\ul}[1]{\underline{#1}}
\nc{\diff}{{\mathcal{T}}}
\nc{\lcdm}[1]{$\Lambda$CDM}
\nc{\Nobs}{N_{\text{obs}}}
\nc{\xoi}{x_{1,i}}

\title[A Bayesian approach to truncated data sets]{A Bayesian approach to truncated data sets: An application to Malmquist bias in Supernova Cosmology}
\author[M.C. March et al.] 
{M.C.~March$^1$\thanks{E-mail: mamarch@sas.upenn.edu},
R.~C.~Wolf$^1$, 
M.~Sako$^1$, 
C.~D'Andrea$^1$, and
D.~Brout$^1$ 
\\ 
$^1$David Rittenhouse Laboratory, University of Pennsylvania, PA, USA
}

\date{Accepted XXX. Received YYY; in original form ZZZ}

\pubyear{2018}

\begin{document}

\label{firstpage}
\pagerange{\pageref{firstpage}--\pageref{lastpage}}
\maketitle

\begin{abstract}
Large scale astronomical surveys are going wider and deeper than ever before.  However, astronomers, cosmologists and theorists continue to face the perennial issue that their data sets are often incomplete in magnitude space and must be carefully treated in order to avoid Malmquist bias, especially in the field of supernova cosmology. Historically, cosmological parameter inference in supernova cosmology was done using $\chi^2$ methodology; however, recent years have seen a rise in the use of Bayesian Hierarchical Models.  In this paper we develop a Bayesian Hierarchical methodology to account for magnitude limited surveys and present a specific application to cosmological parameter inference and model selection in supernova cosmology.    
\end{abstract}


\section{Introduction}\label{sec:intro}

\subsection{Magnitude limited (truncated) astronomical surveys}
A problem commonly encountered in statistical analysis of data is that of truncated data sets.  A truncated data set is one in which a number of data points are completely missing from a sample. This is in contrast to a censored sample in which partial information is missing from some data points. In astronomical observations this problem is commonly seen in a magnitude limited survey such that the survey is incomplete at fainter magnitudes. That is, certain faint objects are simply not observed. The effect of this `missing data' is manifested as Malmquist bias and can result in biases in parameter inference if it is not accounted for.  In Frequentist methodologies the Malmquist bias is often corrected for by analyzing many simulations and computing the appropriate correction factors.  One problem with this methodology is that the corrections are dependent on the model used for the simulations.  In this paper we derive a Bayesian methodology for accounting for truncated data sets  in problems of parameter inference and model selection.  We first show the methodology for a simple Gaussian linear model and then go on to show the method for accounting for a truncated data set in the case for cosmological parameter inference with a  magnitude limited supernova Ia survey.  
\subsection{Evidence for acceleration}
The discovery of the accelerating expansion of the Universe by two independent teams \citep{Riess:1998cb,Perlmutter:1998np} was achieved using a handful of individually named supernovae (SNeIa), meticulous data reduction, and a basic $\chi^2$ minimization method for cosmological parameter inference.  These initial observations and analyses conclusively showed that the Universe is accelerating in its expansion.\footnote{Saul Perlmutter, Brian P. Schmidt and Adam G. Riess were awarded the Noble Prize in physics in 2011 `for the discovery of the accelerating expansion of the Universe through observations of distant supernovae.'} But today we have data from large surveys such as SDSS, SNLS, PanStarrs, DES and others that have yielded hundreds of SNeIa.  With the advent of LSST, we anticipate having thousands of SNeIa. The cosmological questions we are now facing are more nuanced; they seek to understand the cause of acceleration, the nature of dark energy or modified gravity, the isotropy or otherwise of that acceleration.  We wish to know if dark energy evolves with time, whether we live in a \lcdm{} universe or something more exotic.  To address these questions more sophisticated statistical analysis techniques are needed, alongside a better understanding of the systematics. This paper contributes to the ongoing effort to develop increasingly advanced statistical analysis techniques for understanding the subtleties of the physics of the apparent late time acceleration of the Universe.  
\section{Supernova cosmology and Bayesian model selection}
\subsection{The rise of Bayesian methods in supernova cosmology}
The use of Bayesian statistics has become important in supernova cosmology for two reasons: firstly we wish to have a better understanding of the systematics and their effect on the uncertainty in the cosmological parameter estimation, secondly we wish to use Bayesian model selection to discriminate between different theoretical models of dark energy or modified gravity. For an overview of Bayesian methods in cosmology, see e.g., \cite{2008ConPh..49...71T} and \cite{2017arXiv170101467T}.  Some of the early work using SNeIa to do cosmological model selection was done for example by \cite{2000ApJ...530..593D}, \cite{2002PhRvD..65d3506J}, \cite{2004MNRAS.348..603S}, \cite{2004IJMPD..13..669C}, \cite{2005PhRvD..71f3532B}, \cite{2006PhRvD..74l3506L}, and \cite{2006PhRvD..73f7302M}. These applications typically used the basic $\chi^2$ likelihood in their calculation of the SNIa posterior, which meant that the posterior was not correctly normalized and did not account for the uncertainty in the intrinsic dispersion of the SNeIa in a way that was consistent with the Bayesian formalism, nor the degeneracy between the light-curve fitting parameters and the cosmological parameters.  In order to calculate the Bayesian evidence correctly, and to do Bayesian model selection, the posterior must also be calculated in the correct Bayesian manner.

A Supernova Bayesian Hierarchical Model (SN-BHM) to calculate the correctly normalized Bayesian SNIa posterior and model evidence was developed in \cite{2011MNRAS.418.2308M}. This methodology has been implemented in JAGS\footnote{Just Another Gibbs Sampler;\newline \texttt{https://martynplummer.wordpress.com/jags/}} by \cite{2011arXiv1112.3652A} and \cite{2013acna.conf.....H}. 
Other approaches to implementing a Bayesian hierarchical model for SNIa cosmological analysis include \cite{BAMBIS} and Approximate Bayesian Computation (ABC) \cite{2013ApJ...764..116W},\cite{2014PhDT.......156W}, and \cite{2016arXiv161103087J} and was used for various applications \cite{2014MNRAS.437.3298M}, \cite{2013MNRAS.433.2693K}.  Most notably the Bayesian hierarchical model approach was used for analysis of the the UNITY sample \citep{2015ApJ...813..137R} which developed an extended version of the SN-BHM to account for other systematics.  Bayesian hierarchical modeling has also been used in other aspects of SNIa analysis and was first used to determine SNIa properties in the near-infrared in \cite{2009ApJ...704..629M}.
\subsection{Supernova and the selection of exotic cosmological models}
\label{sec:exoticmodels}
Since the development of the Bayesian posterior for SNIa cosmological parameter inference, efforts have continued to attempt to definitively select between different cosmological models, using various forms of Bayesian model selection.  As noted by \cite{2016NatSR...635596N}, some of these methodologies use the full Bayesian Hierarchical Model for the posterior calculation and others use the $\chi^2$ likelihood only. The wealth of papers that continue to be published on this approach, e.g., \cite{2016ApJ...827....1S}, show that there is a definite interest in pursuing a Bayesian approach to discriminating between cosmological models. Examples of cosmological models considered in  Bayesian model selection papers include Chaplygin gas, phantom dark energy, topological defects, dynamical equations of state, brane models, Cardassian models and bouncing models \citep{2006PhLB..633..427S}, Lemaître-Tolman-Bondi models \citep{2010ascl.soft11013G}, Ising perfect fluid \citep{2014IJMPD..2350023L}, dark sector interacting models \citep{2015PhRvD..92j3005W, 2017arXiv171205428F}, DBI\footnote{Dirac-Born-Infeld} actions \citep{2010JPhCS.229a2068C}, timescape models \citep{2012PhDT.......568S}, variable dark energy \citep{2007ChPhL..24.2459X}, dynamical dark energy \citep{2012PhRvL.109q1301Z}, two field inflation models \citep{2012ApJ...753..151V}, adiabatic LTB models \citep{2012JCAP...10..009Z}, ghost free bi-gravity \citep{2013JHEP...03..099A}, Brans-Dicke cosmologies \citep{2014PhRvD..90l4040H}, Bianchi spacetimes \citep{2017arXiv171202072A}, and holographic dark energy \citep{2017arXiv171002417M}. Another avenue for exploration with Bayesian model selection has been to test the internal or external consistency of different data sets (e.g., \citet{2006PhRvD..73f7302M}, \citet{2014MNRAS.439.1855H}, \citet{2015MNRAS.449.2405K}). Bayesian model selection is rapidly becoming the statistical method of choice in the quest to determine what type of universe we inhabit, hence it is worth investing in the development of the overall Bayesian framework for SNIa cosmological analysis.
\subsection{Current limits and the future of model selection with SN}
Thus far the data have been insufficiently constraining to provide strong Bayesian evidence for a particular cosmological model, according to the Jeffreys scale \citep{Jeffreys:1961}. However now in the era of stage III dark energy projects,\footnote{As defined by \cite{2006astro.ph..9591A}.} (e.g., PanStars \citep{2016arXiv161205560C} and the Dark Energy Survey \citep{2005IJMPA..20.3121F}, each expected to have several thousand SNeIa on completion) with the advent of stage IV ground based surveys such as LSST which will increase the sheer number of $z<1$ SN to tens of thousands \citep{2009arXiv0912.0201L}, and stage IV space based missions such as WFIRST \cite{2015arXiv150303757S} which will increase the redshift range of the SN sample to $z\sim2$, we will be able to considerably improve model constraints. 

In order for Bayesian model selection to be an effective tool, the systematics of the survey which appear in the posterior probability must also be treated in a robust Bayesian way, for example, selection effects, which are the focus of this paper. Stage III and stage IV surveys will face additional systematics challenges:  the very large samples will be photometric samples (lacking spectroscopic classifications or redshifts) rather than spectroscopic samples which do have spectroscopic classifications and redshifts.  Work has been done in treating these systematics in a Bayesian framework by \cite{2015mgm..conf.2069K} and \cite{2017JCAP...10..036R}, which are other important contributions to the development of a full Bayesian methodology for a photometric sample. 
\section{Supernova cosmology and Malmquist bias}
In cosmological applications, SNeIa are used as standardisable candles for parameter inference and model selection. The key observables are the redshift and variation of magnitude in time and color space (i.e. the light curve), which are summarized by the peak B-band magnitude in the observed frame, $m_B$, the color parameter, $c$ and the light curve shape parameter $x_1$ for each observed supernova.  The cosmological parameters can be inferred by comparing the theoretical and `observed' distance moduli of each SN, as related through the SALT-II equation. The SALT-II equation which is used to standardize the SNeIa is as follows
\begin{align}
\label{eq:salt}
\mu_i = \mbi - M_0 + \alpha \xoi - \beta c_i +\epsilon_i \, ,
\end{align}
where $\alpha$ and $\beta$ are unknown and must be inferred from the data simultaneously with the cosmological parameters. $\epsilon_i$ represents the intrinsic dispersion in the absolute magnitude, $M_0$ of the SNeIa and is modeled as being drawn from a Gaussian distribution
\be
\epsilon_i \sim \mathcal{N}(0,\sigint) \, .
\ee
\subsection{Cosmology from Supernovae Type Ia}
In terms of the cosmological parameters $\Cp$ (which includes the matter density $\Om$, dark energy density $\OL$, dark energy equation of state $w$, and Hubble parameter $H_0$), the distance modulus for the $i$th SNIa at redshift $z_i$ is 
\begin{align} \label{eq:mucosmo}
\mu_i &= \mu(z_i, \Cp) \\ &= 5 \log \left[ \frac{D_L(z_i, \Cp)}{\Mpc} \right] +25 \, ,
\end{align}
where $\Cp$ denotes the cosmological parameters
\be
\Cp = \{\Om, \OL, w, H_0 \} \, ,
\ee
and where the curvature density $\Omk$ is constrained by
\be
\Omk = 1 - \Om - \OL \, .
\ee
The luminosity distance $D_L =\frac{c}{H_0}d_L$ is given in terms of the dimensionless luminosity distance $d_L$
\begin{align}
d_L(z,\Om, \OL, w) = \frac{(1+z)}{\sqrt{|\Omk|}}\text{sinn}& \{ \sqrt{|\Omk|}  \int_0^z \dr z' 
\left[ (1+z')^3\Om \right. \nonumber \\ 
 \left. + \Ode(z') + \right.  \left. (1+z')^2\Omk \right]^{-1/2} \} \, ,
\end{align}
assuming a negligible contribution from the radiation density, where $\text{sinn}(x) = \sinh(x)$ if $\Omk \le 0$ and $\text{sinn}(x) = \sin(x)$ if $\Omk \ge 0.$
The dark energy density parameter $\Ode(z) $ is given by
\be
\Ode(z) =  \OL \exp\left(3\int_0^z \frac{1+w(x)}{1+x}\dr x \right) \, .
\ee
\subsection{Malmquist Bias}
\begin{figure}
\centering
\includegraphics[width=1.0 \linewidth]{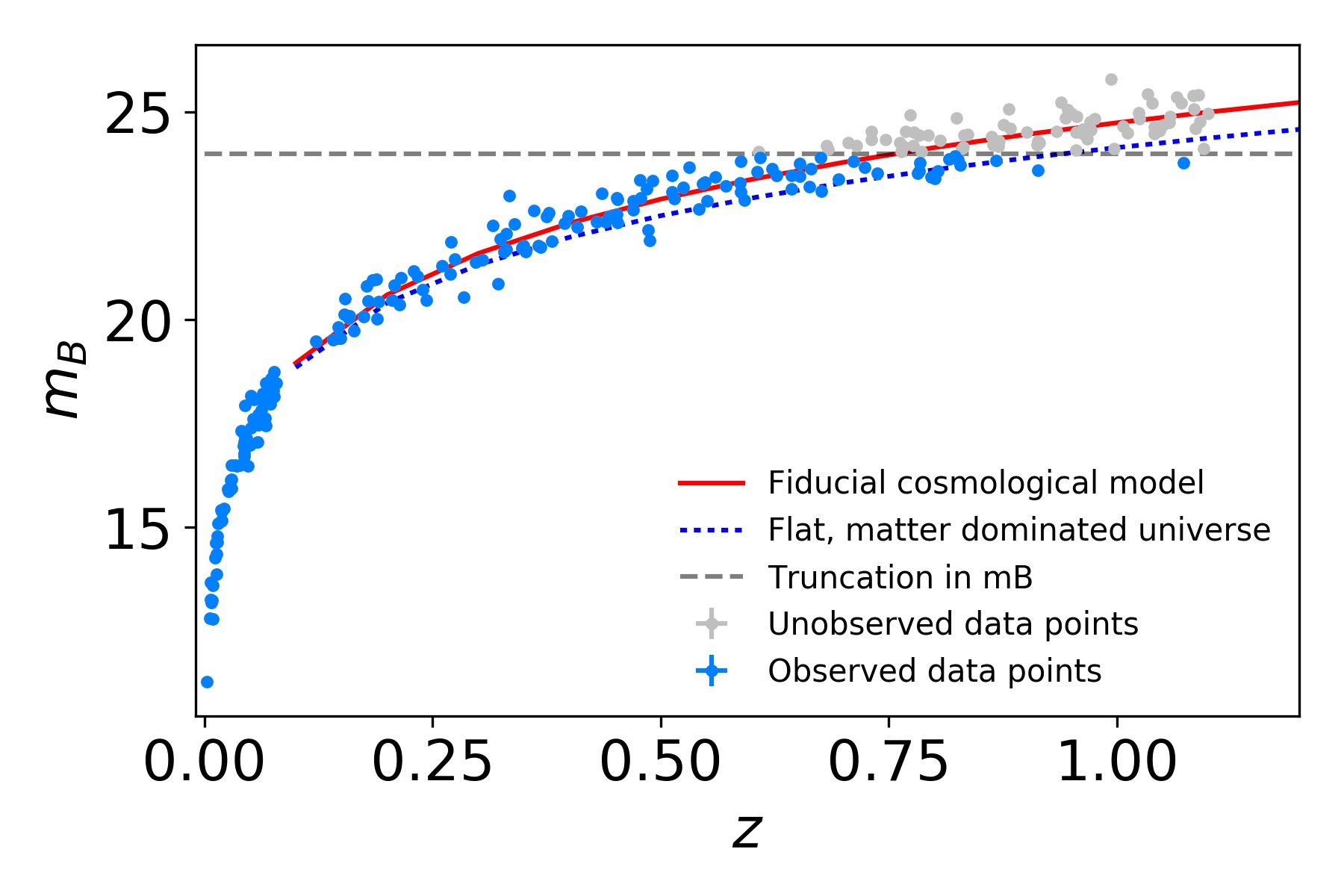} 
\caption{A simulated data set of 250 supernova. A cut is imposed such that data points with a peak B-band magnitude $mB>24.0$ are discarded leaving a truncated data set. The solid red line shows the fiducial model.  The dotted blue line shows that a matter dominated model would be incorrectly favored if the truncated data set alone were considered without explicitly accounting for the truncation. This is the effect of Malmquist bias.}
\label{fig:cosmo_data_scatter_malmquist}
\end{figure}

The inference problem we face in a magnitude limited SNIa survey is that due to the sensitivity limit of the telescope camera, and environmental factors, supernova fainter than a certain magnitude are missing from the observed sample, leaving us with a truncated data set. See Fig.~\ref{fig:cosmo_data_scatter_malmquist} for an example of a model truncated SN data set. Using this truncated data set to do cosmological parameter inference without correction would result in a bias known as Malmquist bias. This problem has historically been dealt with by 
modeling the effect of Malmquist bias through simulations, then applying a correction factor based on the simulations, and adding the uncertainty to the systematic error budget \citep{Riess:1998cb,Perlmutter:1998np,Astier:2005qq,2009ApJS..185...32K}. A limitation of this method is that corrections are made for missing data points by adjusting the magnitude or luminosity distance of \emph{existing} data points, rather than accounting for the \emph{missing} data points. A more consistent method is to apply the corrections to binned data, such that corrections are made to compensate for biases over a redshift bin.  For examples of this methodology where corrections are applied to binned data, see \cite{2017ApJ...836...56K} and \cite{2017arXiv171000845S}.  However, all of these examples depend on corrections based on simulations which assume a particular underlying cosmological model, such as $\Lambda$CDM or $w$CDM. It is not clear that it is consistent to use these corrections when considering parameter inference for some of the more exotic cosmological models mentioned in section \ref{sec:exoticmodels}, or indeed model selection between such models.  Another approach could be to only use the part of the data which is complete in magnitude space.  This would result in an unbiased sample, but there would be a loss of information as higher redshift data is discarded.

This paper seeks to develop a method that does not require simulations or loss of information through discarding data, using a Bayesian framework.  The current most advanced methodology for doing so is \cite{2015ApJ...813..137R}, which presents a very detailed Bayesian analysis which accounts for a number of systematic effects. However it does not derive an exact solution for Malmquist bias, but uses an approximation which breaks down in severe cases of Malmquist bias. The solution derived in this paper could replace the approximation mentioned.

The problem of cosmological parameter inference from SNeIa data fitted using the SALT-II light-curve model is essentially a variation on regression in a Gaussian linear model, which indeed forms the basis of many inference problems using astronomical data.  We will proceed by first solving the general problem of inference in a truncated data set in a multidimensional Gaussian model, then proceed to apply it to the special case of cosmological parameter inference in a magnitude limited SNIa survey.

\section{Truncated data sets in the Gaussian linear model}\label{sec:glam}

We begin by considering the case of a Gaussian linear model, since this forms the basis of the model we shall use for the SNIa cosmology case. This model also forms the basis of many other parameter inference problems in astronomy, and we hope this methodology will be useful for those working on similar problems other areas of research. 

\subsection{Bayes equation for Gaussian linear model}
\label{sec:glmbayes}
Let us consider a Gaussian linear model with a vector of $J$ independent variables $\bf{x}_j=[x_1, \dots x_j, \dots x_J]$, a dependent variable $y$ subject to a small intrinsic dispersion $\epsilon$ with a J-dimensional vector of slope parameters $\bf{a}=[a_1, \dots a_j, \dots a_J]$ and intercept parameter $b$ which can be written as
\be
y_i = \mathbf{ax}_i +b + \epsilon_i \, ,
\label{eq:glmtrue}
\ee
where the intrinsic dispersion is characterized by
\be
\normal {\epsilon_ i} {0} \sint \, .
\ee
The relationship in Eq.~\ref{eq:glmtrue} is exact and the variables $\bf{x_i}$ and $y_i$ are the true or latent variables which are never seen.  Instead, we measure the observed variables  $\bf{\xhi}$ and $\yhi$ which are subject to observational noise characterized by
\begin{align}
& \normal {\mathbf{\xhi}}{\mathbf{x}_i}{\mathbf{\sxi}} \, , \\
& \normal \yhi {y_i} \syi   \, ,
\end{align}
where
\be
\bf{\sxi} \in \mathbb{R}^{J \times J} \, .
\ee
Two questions of interest given a data set $(\bf{\xh}, \yh)$ of all the observed $\bf{x}_i$ and $y_i$ are firstly the inference of the slope and intercept parameters $\bf{a}$ and $b$ and secondly the likelihood that the model, $M$ is indeed the best model given the data.  The answers to these two questions can be obtained from Bayes' equation
\be
\label{eq:Bayes}
p ( \mathbf{a},b | \mathbf{\xh}, \yh, M ) = \frac{p  ( \mathbf{\xh}, \yh | \mathbf{a},b, M)  p ( \mathbf{a},b | M ) }{p (\mathbf{\xh}, \yh | M )} \, .
\ee 
The probability of $\bf{a}$ and $b$ given the observed data is given by the posterior $p (\mathbf{a},b | \mathbf{\xh}, \yh , M )$ and the model likelihood is given by the Bayesian evidence $p (\mathbf{\xh}, \yh | M )$.  The derivation of the posterior probability of $\bf{a}$ and $b$ given the observed data is shown in Appendix \ref{sec:glmlike}.  The posterior described in Eq.~\ref{eq:GLM1post} is the same as that given by \cite{Gull1989} and \cite{2011MNRAS.418.2308M} re-cast into the formalism of \cite{2007ApJ...665.1489K}.  The aforementioned papers have shown that this likelihood can be inserted into Eq.~\ref{eq:Bayes} and used to give unbiased point estimators for the slope and intercept parameters in challenging cases where there are large error bars on both the dependent and independent variables.  

But suppose we are trying to infer parameters from a data set in which part of the data are missing or truncated. This is the key question addressed in this paper, for example as shown in Fig.~\ref{fig:glm_data_scatter_1}. Fitting a Gaussian linear model to the truncated data alone without accounting for the truncation would result in an underestimate of the gradients of the slopes. The posterior must be expanded with inclusion terms to account for the uncertainty about whether a particular data point is included or excluded from the observed data set (i.e., in Eq.~\ref{eq:I1} and \ref{eq:I2}).

Adapting the methodology of \cite{BDA} and \cite{2008ApJ...682..874K}, we derive a Bayesian expression for the unbiased estimator of the parameters of interest in the case where part of the data are missing. We consider the case where the data are not missing at random, but rather are missing based on some condition.  We consider the case where data points are missing if their value exceeds some threshold in the dependent variable, $y^{\text{thresh}}$.   

Let us consider the case where a theoretical complete data set of $N$ data points could have been measured, but out of those $N$ potential measurements, information is missing for $m$ data points such that there are only $\Nobs$ observed data points with in the observed data set.  If $N$ is known the data set is said to be censored, since at least the \emph{number} of missing data points is known. If $N$ is unknown, the data set is said to be truncated.  The number of missing data points is given by
\be
m=N-\Nobs \, .
\ee
In the truncated case we consider that data points with a $y$ value greater than some threshold  $y^{\text{thresh}}$ are missing from the observed data set.  The inclusion model $I$ is such that $I = 1$ for data included in the observed data set and $I = 0 $ for data missing from the observed data set, i.e.,
\begin{align}
\label{eq:I1}
p(I_i =1 | \yhi) &= 1 \,\, \text{if} \,\,\, \yhi \le  y^{\text{thresh}}  \\
p(I_i =0 | \yhi) &= 1 \,\, \text{if} \,\,\, \yhi >  y^{\text{thresh}} \, .
\label{eq:I2}
\end{align}
To find the probability of the observed data points, we need to marginalize over the missing data points. The details of this calculation showing how to include the additional uncertainty in the posterior is shown in Appendix \ref{sec:glmtrunc}, with the final result being given by Eq.~\ref{eq:GLM1postI}.

\subsection{Numerical tests of parameter inference for Gaussian linear model with truncated data set}
\begin{figure*}
\centering
\includegraphics[width=0.3 \linewidth]{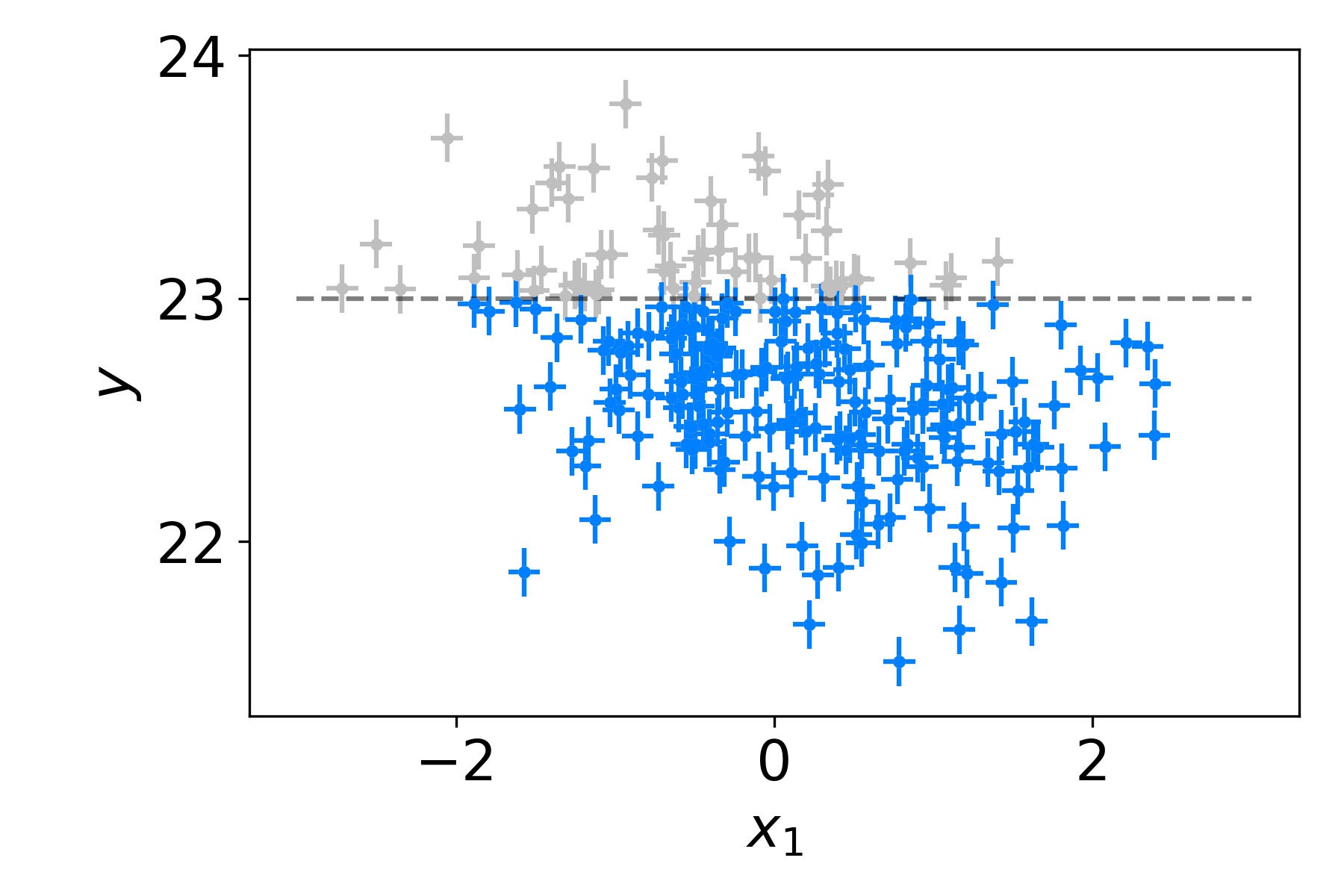} \hfill
\includegraphics[width=0.3 \linewidth]{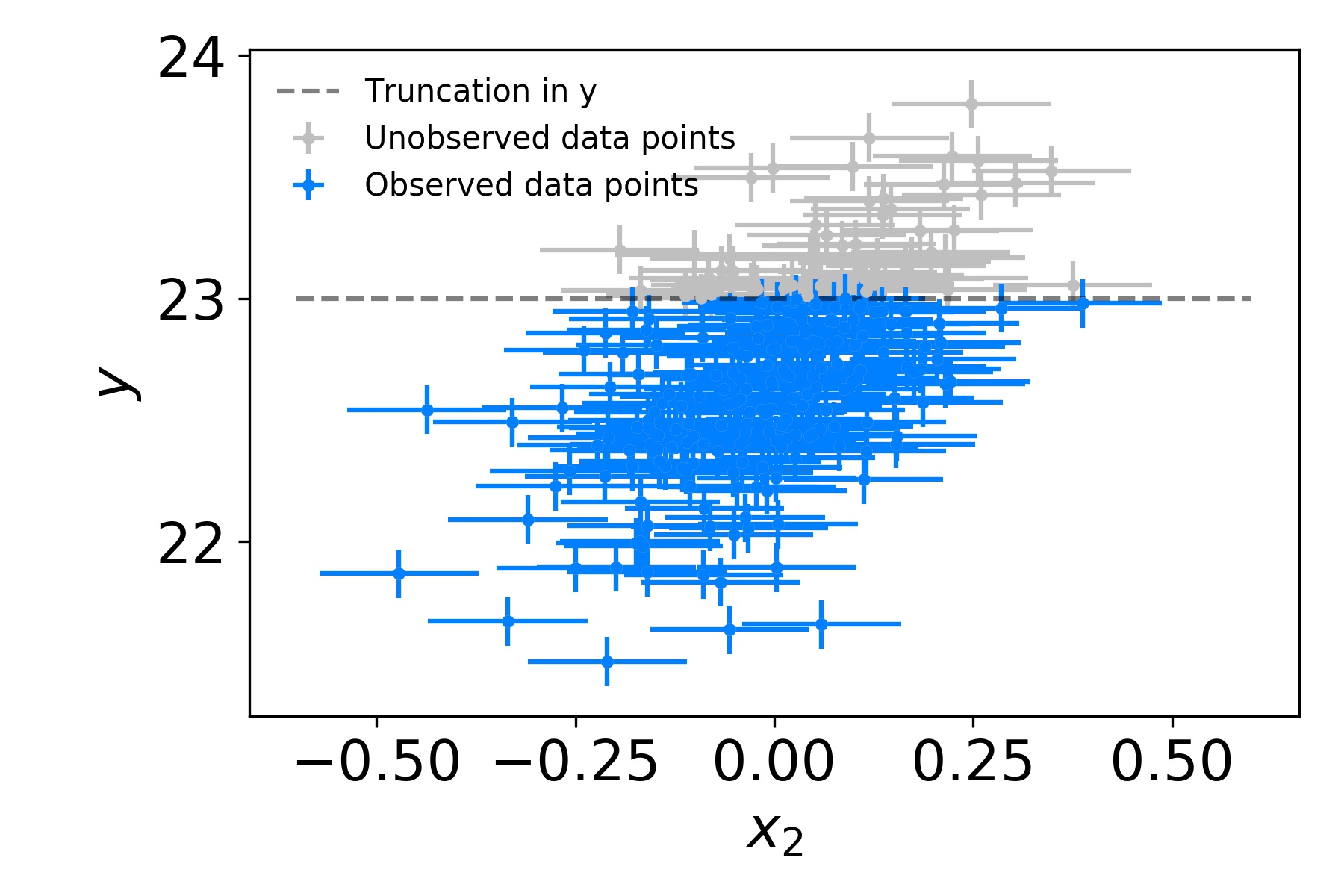} \hfill
\includegraphics[width=0.3 \linewidth]{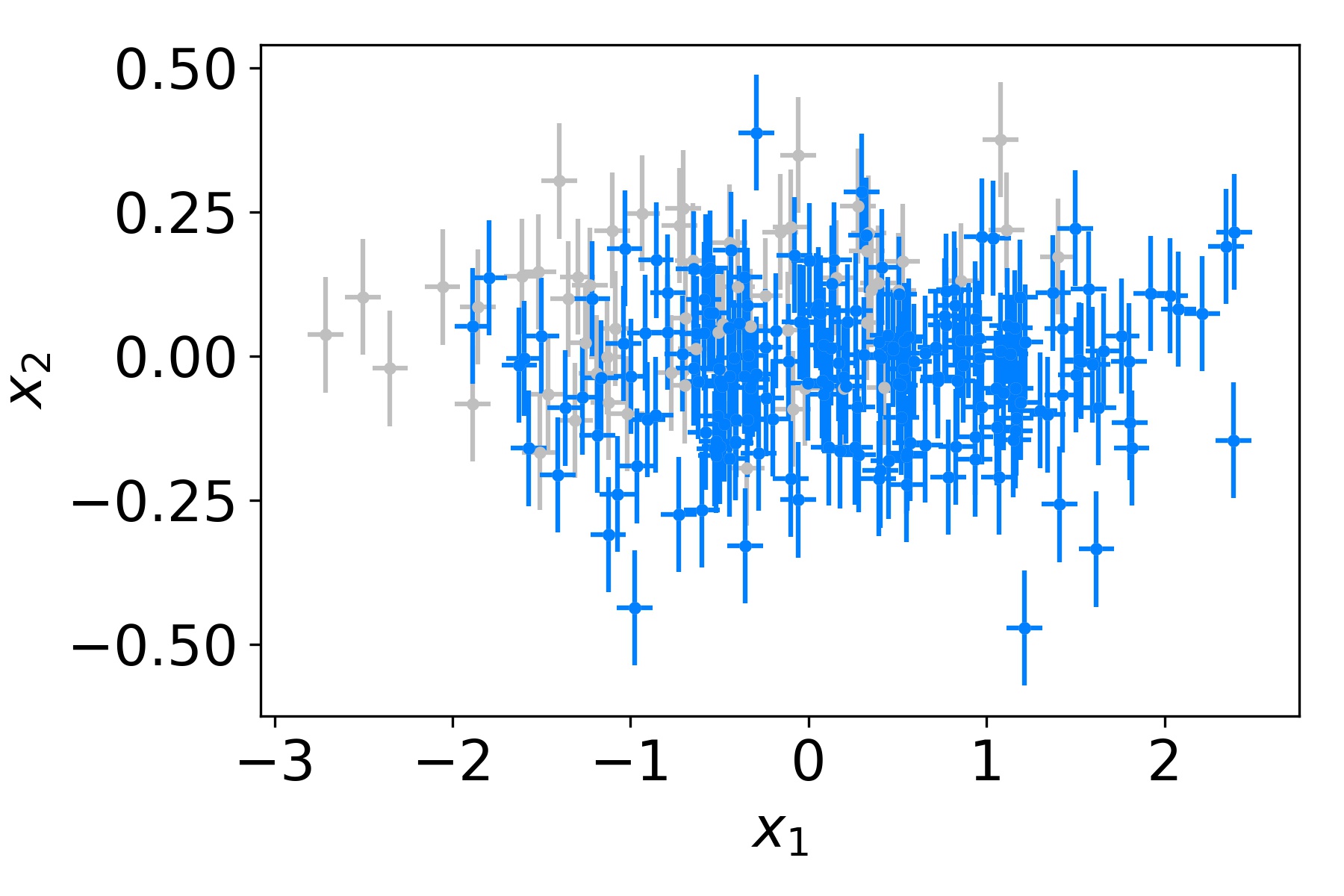} 
\caption{A simulated data set for $J=2$ Gaussian linear model.  Blue and grey points together make up the full data set.  Blue points are observed. Data points with $y>23.0$ are discarded (grey), leaving a truncated data set (blue points).}
\label{fig:glm_data_scatter_1}
\end{figure*}

Let us consider the example of a Gaussian linear model with two dependent variables, i.e. $J = 2$ in Eq.~\ref{eq:glmtrue}, such that we have
\be
\label{eq:glmJ2}
y_i = a_1x_{1,i} + a_2x_{2,i} + b + \epsilon_i \, .
\ee
We simulated data sets with the characteristics described in Table \ref{tab:simglm} and chose to make a cut at $y^{\text{obs}}_i >23.0$, discarding all data that fell outside of that limit.  This cut-off was chosen such that approximately $25\%$ of the data set was discarded or truncated. 100 sets of data were simulated each having 250 data points (these are the complete data sets).  For each complete data set there is a corresponding truncated data set for which the $y^{\text{obs}}_i >23.0$ cut had been applied. Examples of the complete and truncated data sets are shown in Fig.~\ref{fig:glm_data_scatter_1}. The posterior probability distribution for the truncated data set, given in Eq.~\ref{eq:GLM1postI} was sampled using the Multinest sampling algorithm and associated code \citep{2009MNRAS.398.1601F} with the pyMultinest wrapper \citep{2014A&A...564A.125B}. Resulting samples were processed using Chain Consumer \citep{Hinton2016}.

\begin{figure*}
\centering
\includegraphics[width=1.0 \linewidth]{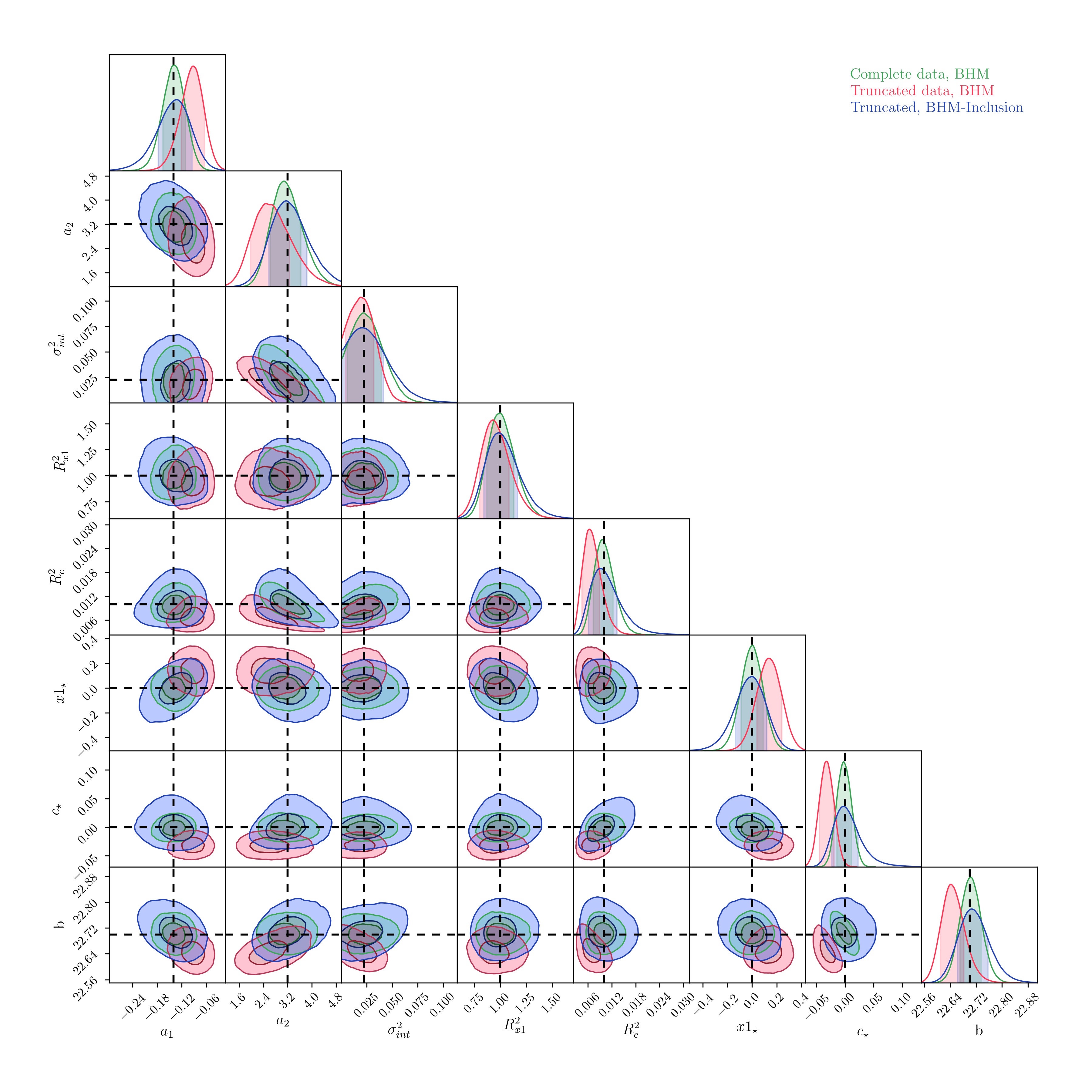} 
\caption{Parameter inference for 2-dim Gaussian linear model. Inner and outer contours enclose $68.3\%$ and $95\%$ of the posterior mass respectively. Green contours show complete (observed and unobserved) data set fitted with the standard Bayesian hierarchical model. Red contours show the truncated data sets fitted with the basic Bayesian hierarchical model, blue contours show the truncated data sets fitted with the extended Bayesian hierarchical model with the inclusion model, that accounts for the truncation.}
\label{fig:glm_tri}
\end{figure*}

The results in Fig.~\ref{fig:glm_tri} show that when the posterior for the basic Bayesian hierarchical model described in \cite{2011MNRAS.418.2308M} and summarized by Eq.~\ref{eq:GLM1post} is used with the complete data set, all parameters are correctly inferred.  When the truncated data sets are analyzed using that basic SN-BHM method, we see expected biases in the recovery of the parameters of interest, most notably the slopes of the gradients are underestimated.  However, when the truncated data sets are analyzed using the Bayesian hierarchical model with the inclusion model, as described in Eq.~\ref{eq:GLM1postI} the parameters are accurately inferred, with slightly large uncertainties.  The larger interval estimators are expected and reflect the fact that there is increased uncertainty due to marginalizing over missing data points.

\begin{table}
\centering
\begin{tabular}{lll}
\hline\hline 
Parameter & Symbol & True Value   \\ \hline 
Intercept & $b$ & 22.7  \\
Gradient 1 & $a_1$& -0.14 \\
Gradient 2  & $a_2$ & 3.2 \\ \hline
Mean of distribution of  $\underline{x}_1$ & $\xnot$ & 0.0 \\
Mean of distribution of  $\underline{x}_2$ & $\cnot$ & 0.0 \\
s.d. of distribution of  $\underline{x}_1$ & $\Rx$ & 1.0 \\
s.d. of distribution of $\underline{x}_2$ & $\Rc$ & 0.1 \\
Observational noise on $\underline{y}$ & $\sigma_{y}$ & 0.1\\
Observational noise on $\underline{x}_1$ & $\sigma_{x_1i}$ & 0.1 \\
Observational noise on $\underline{x}_2$ & $\sigma_{ci}$ & 0.1 \\
\hline
\end{tabular}
\caption{Input parameter values used for the fiducial model in the generation of the simulated $J=2$ Gaussian linear model data sets.}
\label{tab:simglm}
\end{table}

\section{Truncated data sets in supernova cosmology}

\subsection{Posterior for supernova data when truncated in Magnitude space}

The SALT-II equation takes the form of a bivariate Gaussian linear model described in Eq.\ref{eq:glmJ2} with the `intercept' parameter $b$ being 
\onecolumn
\noindent
replaced by the sum of a cosmology and redshift dependent function $\mu(\Cp,z_i)$ defined in Eq.~\ref{eq:mucosmo} and the intrinsic magnitude $M_0$. We reproduce Eqs.\ref{eq:glmJ2} and \ref{eq:salt} below for comparison
\begin{align}
y_i &= a_1x_{1,i} + a_2x_{2,i} + b + \epsilon_i   \tag{\ref{eq:glmJ2}} \, ,\\ 
m_{B,i} &= \alpha x_{1,i} - \beta c_i + (\mu(\Cp,z_i) +M_0 ) + \epsilon_i \tag{\ref{eq:salt}} \, .
\end{align}
 GLM sims data: scatter
\noindent
The measured light curve fit parameters and covariance matrices are denoted by
\begin{align}
\mathbf{\whi} &= \begin{bmatrix} 
\hat{m}_{B,i}  \\ 
\hat{x}_{1,i}    \\
\hat{c}_i \end{bmatrix} \in \mathbb{R}^3 \, ,\\ 
\Sigma_{C,i} &=
\begin{bmatrix} 
\sigma_{mi}^2 &  \sigma_{mi, x1i} & \sigma_{mi, ci}\\
\sigma_{mi,x1i} & \sigma_{x1i}^2 &  \sigma_{c,x1i} \\ 
\sigma_{mi,ci} & \sigma_{x1i,ci} &  \sigma_{ci}^2 \\ 
\end{bmatrix}  \in \mathbb{R}^{3 \times 3} \, .\\
\end{align}
The means and standard deviations of distributions of the latent variables $c_i$ and $x_{1,i}$ are:
\begin{align}
\mathbf{\xnot} &= \left[ x_{1,\star} c_{\star}\right] \in \mathbb{R}^2 \nonumber \, , \\
\mathbf{\Rx} &= \begin{bmatrix} R_{x1} & 0 \\ 0 & R_c \end{bmatrix} \in  \mathbb{R}^{2 \times 2} \, . \\ 
\end{align}
The redshift uncertainties are
\be
\hat{z}_i \sim \norm \left( z_i,\sigma_z \right) \, .
\ee 
Following \cite{2011MNRAS.418.2308M}, after integrating over the latent $z_i$, the redshift uncertainty can be included as a contribution to $\sigma_{m_{Bi}}$ such that  $\sigma_{m_{Bi}} \to \sigma_{m_{Bi}}^{\text{raw}} + f_i\sigma_{zi}f_i$ where
\be
f_i = \log_{10}(e)\left.\frac{D_L'(z_i)}{D_L(z_i)}\right|_{\hat{z}_i} \, ,
\ee 
and $\sigma_{m_{Bi}}^{\text{raw}}$ denotes the original error on $m_{Bi}$. The posterior probability of the cosmological and SALT-II parameters, given the SNIa data is of a similar form to Eq.~\ref{eq:GLM1like} with $J=2$ such that
\begin{align}
p ( \Cp,\alpha,\beta |x_1^{\text{obs}},c^{\text{obs}} m_B^{\text{obs}},z^{\text{obs}},m_B^{\text{thresh}}I,M )
&\propto \int_{\Nobs}^{\inf} dN \, \iint   \, d\mathbf{\Rx} \, d\mathbf{\xnot} \, \frac{1}{N} \binom{N}{\Nobs} \prod_i^{\Nobs}   | 2\pi \Sigma^{\text{obs}} _{v,i}|^{-\frac{1}{2}} \exp{ \left( -\frac{1}{2} \left( (\mathbf{\whi}^{\text{obs}} -\mathbf{q}_i)^T \Sigma^{\text{obs}^{-1}} _{v,i} (\mathbf{\whi}^{\text{obs}} -\mathbf{q}_i )\right)\right)} \nonumber \\
& \times \prod_i^{m} \int_{-\infty}^{+\infty} \int_{-\infty}^{+\infty} \int_{m_B^{\text{thresh}}}^{+\infty}   \int_{0}^{\hat{z}^{\text{max}}}d\hat{x}_{1i}^{\text{mis}} d\hat{c}_i^{\text{mis}} d\hat{m}_{Bi}^{\text{mis}}\, d\hat{z}_i^{\text{mis}}  | 2\pi \Sigma^{\text{mis}} _{v,i}|^{-\frac{1}{2}} \nonumber \\
&\times \exp{ \left( -\frac{1}{2} \left( (\mathbf{\whi}^{\text{mis}} -\mathbf{q}_i)^T \Sigma^{\text{mis}^{-1}} _{v,i} (\mathbf{\whi}^{\text{mis}} -\mathbf{q}_i )\right)\right)} \nonumber \\ 
&\times p(\Rx, \xnot |I,M) p(\Cp,\alpha, \beta |I,M) \, . \label{eq:SNBHMt1post}
\end{align}
Here the highest observed redshift has been chosen as a reasonable upper limit for the redshift integral. The remaining terms in the likelihood contribution to the posterior are
\begin{align}
\mathbf{q}_i &= \begin{bmatrix} 
\mu(\Cp,\hat{z}_i) +M_0  + \alpha x_{1,\star}  - \beta c_{\star}  \\ 
x_{1,\star} \\
c_{\star}
\end{bmatrix}  \in \mathbb{R}^3 \, , \\
\Sigma_{v,i} &=
\begin{bmatrix} 
\sigma_{mi}^2 + \sint^2  + \alpha^2R_{x1}^2 + \beta^2R_{c}^2, &  \sigma_{mi, x1i} + \alpha R_{x1}^2,  & \sigma_{mi, ci} + \beta R_{c}^2  \\
\sigma_{mi,x1i} + \alpha R_{x1}^2,  & \sigma_{x1i}^2 + R_{x1}^2, &  \sigma_{c,x1i} \\ 
\sigma_{mi,ci} +  \beta R_{c}^2, & \sigma_{x1i,ci}, &  \sigma_{ci}^2  + R_{c}^2  \\ 
\end{bmatrix}  \in \mathbb{R}^{3 \times 3} \, . \\
\end{align}
This is the principle result of this paper.  It is a Bayesian derivation of the posterior probability of the cosmological parameters given the SNIa data for a truncated data set, such as are to be found in magnitude limited surveys.  This posterior accounts for Malmquist bias by using an inclusion model which describes the truncation in magnitude space. 
%

\begin{table}
\centering
\begin{tabular}{lll}
\hline\hline 
Parameter & Symbol & True Value   \\ \hline 
Matter energy density parameter&$\Om$ & 0.3  \\
Dark energy density parameter &$\OL$ & 0.7  \\
Dark energy equation of state & $w$ & $-1$ \\
Spatial curvature & $\Omk$ & 0.0 \\ 
Hubble expansion rate &$H_0$ [km/s/Mpc] & 72.0 \\\hline
Mean absolute magnitude of SNe & $M_0$ [mag]  & -19.3   \\
Intrinsic dispersion of SNe magnitude& $\sigint$ [mag] & 0.1 \\ 
Stretch SALT II parameter & $\alpha$& -0.14 \\
Color SALT II parameter  & $\beta$ & 3.2 \\ \hline
Mean of distribution of  $\xbl$ & $\xnot$ & 0.0 \\
Mean of distribution of  $\cbl$ & $\cnot$ & 0.0 \\
s.d. of distribution of  $\xbl$ & $\Rx$ & 1.0 \\
s.d. of distribution of $\cbl$ & $\Rc$ & 0.1 \\
Observational noise on $\mb$ & $\sigma_{\mbi}$ & 0.1\\
Observational noise on $\xbl$ & $\sigma_{x_1i}$ & 0.1 \\
Observational noise on $\cbl$ & $\sigma_{ci}$ & 0.1 \\
Correlation between $\xbl$ and $\cbl$ & $\sigma_{x_1i,ci}$ & 0.0 \\
\hline
\end{tabular}
\caption{Input parameter values used for the fiducial model in the generation of the simulated SNe \salt{} data sets.}
\label{tab:simsn}
\end{table}
%
\section{Numerical trials with simulated supernova data}

\begin{figure*}
\centering
\includegraphics[width=0.3 \linewidth]{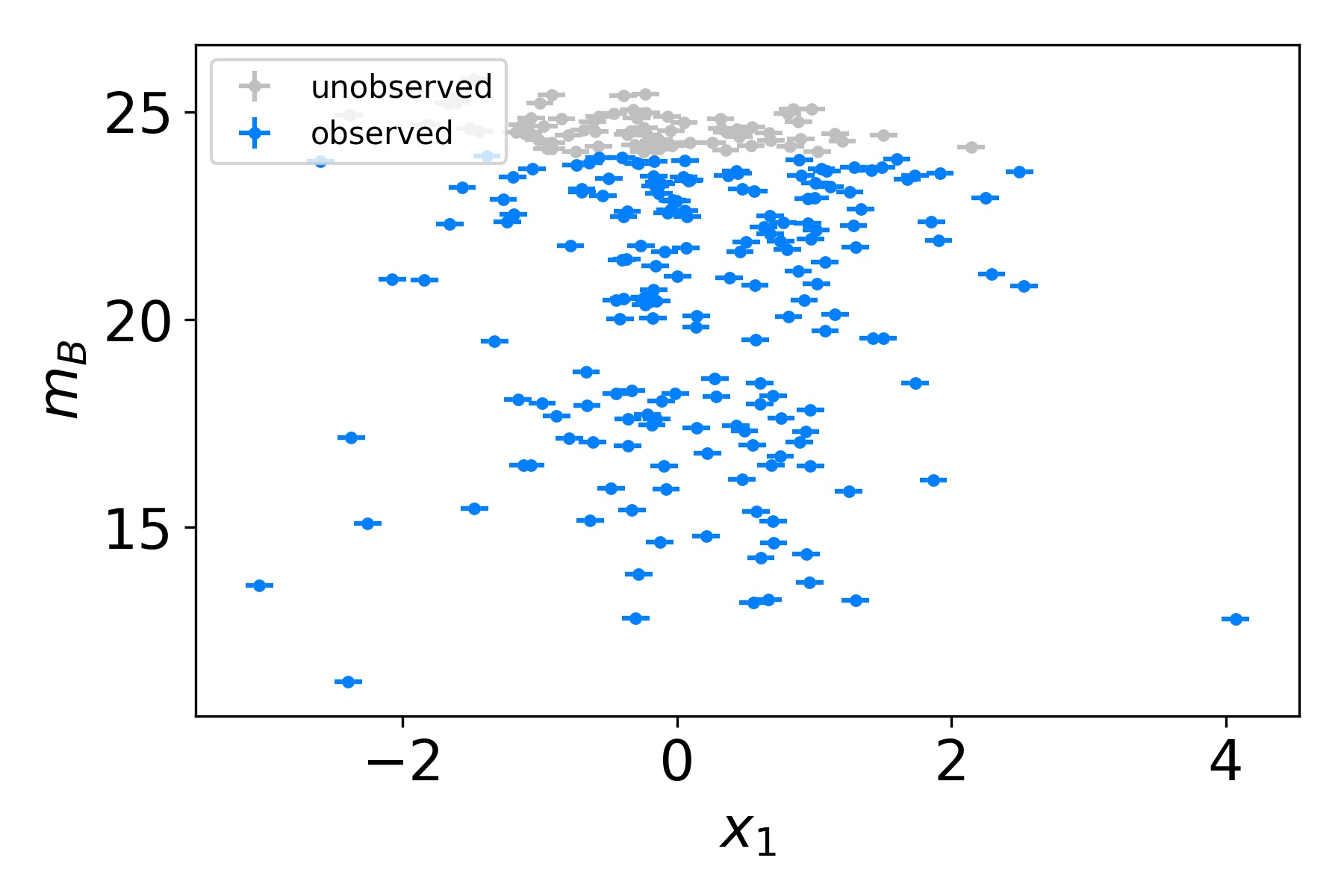} \hfill
\includegraphics[width=0.3 \linewidth]{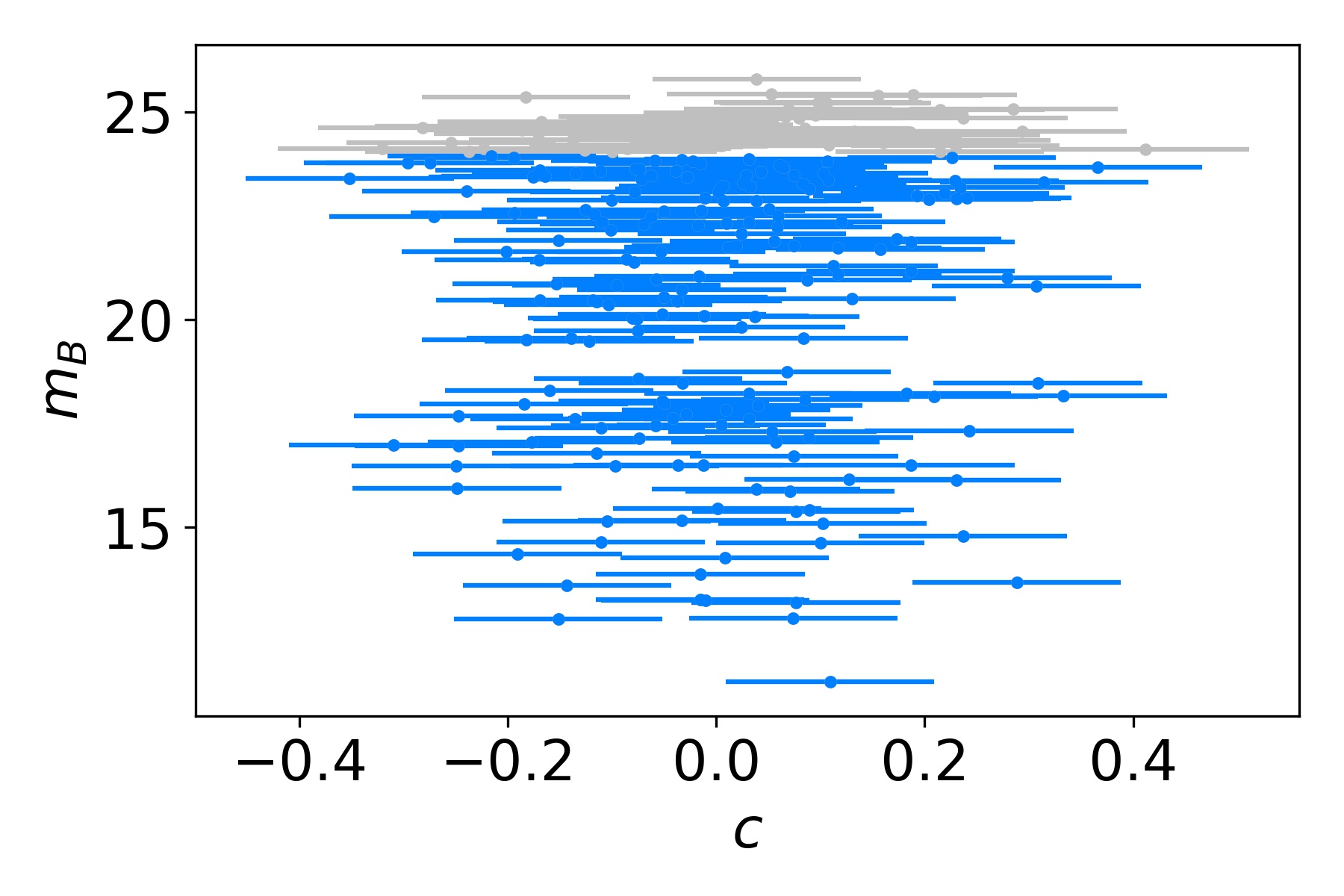} \hfill
\includegraphics[width=0.3 \linewidth]{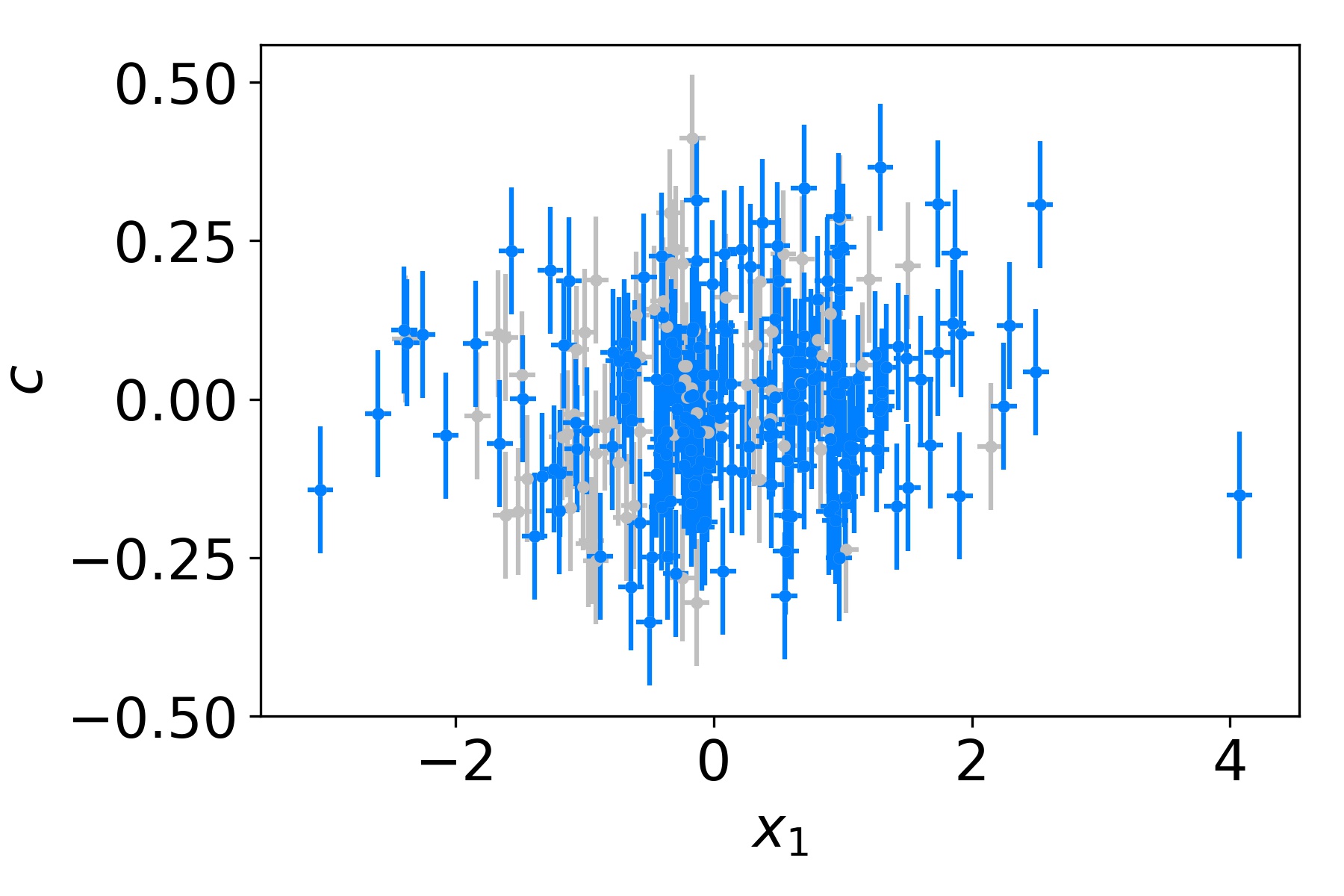} 
\caption{A simulated data set using the SALT-II supernova model.  Blue and grey points together make up the full data set.  Blue points are observed, grey points are not observed and the data set is said to be truncated.}
\label{fig:cosmo_data_scatter_1}
\end{figure*}

\begin{figure*}
\centering
\includegraphics[width=0.3 \linewidth]{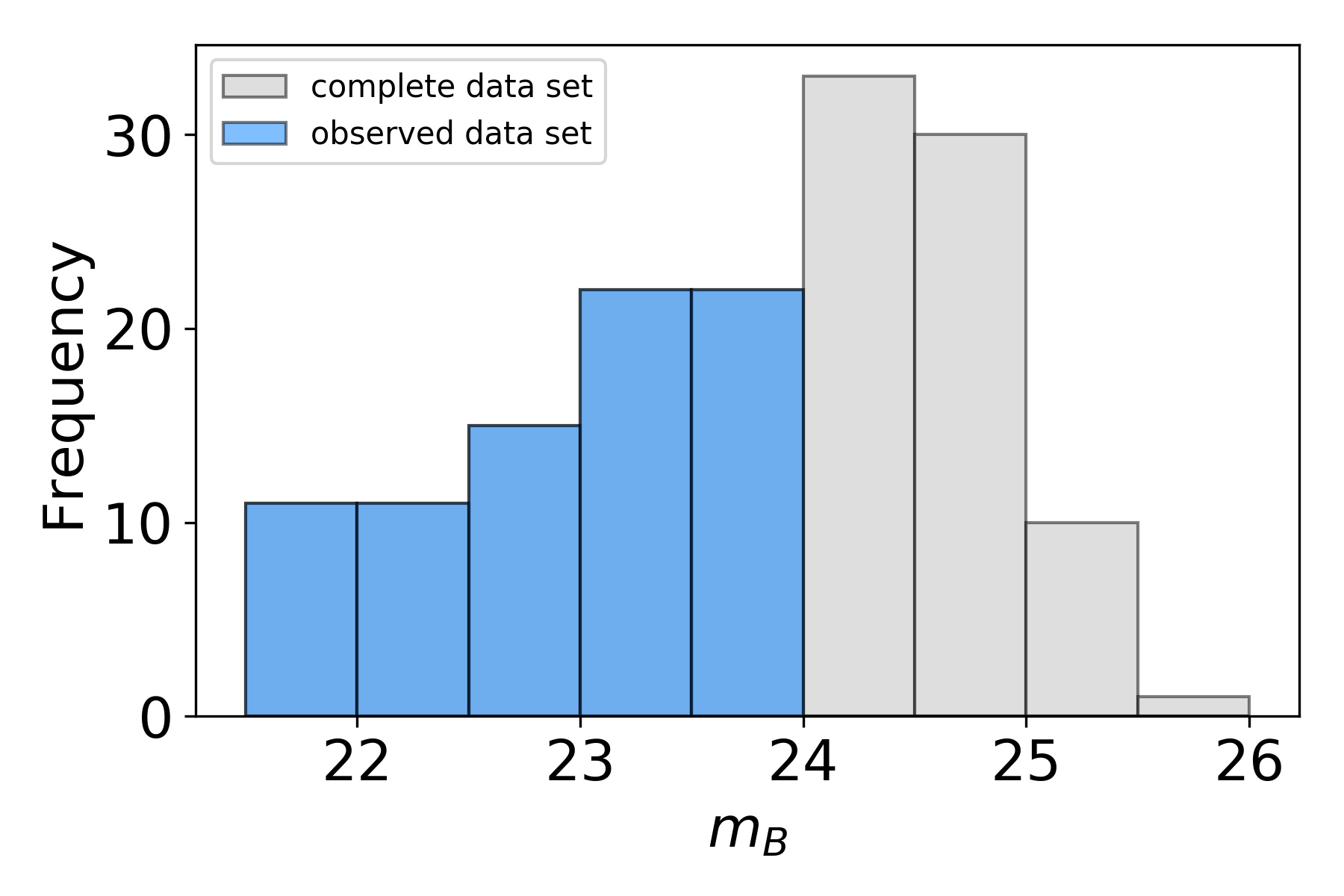} \hfill
\includegraphics[width=0.3 \linewidth]{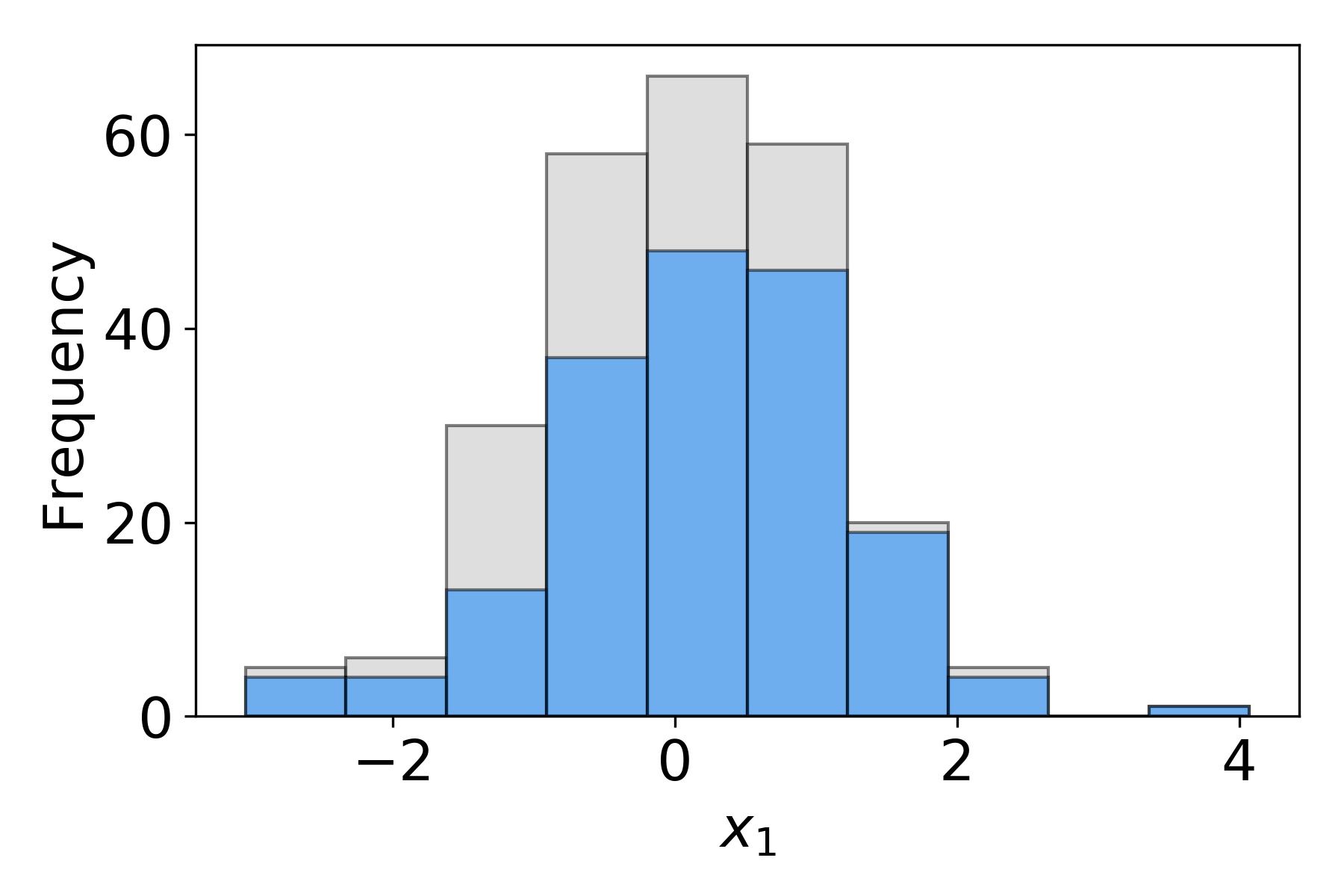} \hfill
\includegraphics[width=0.3 \linewidth]{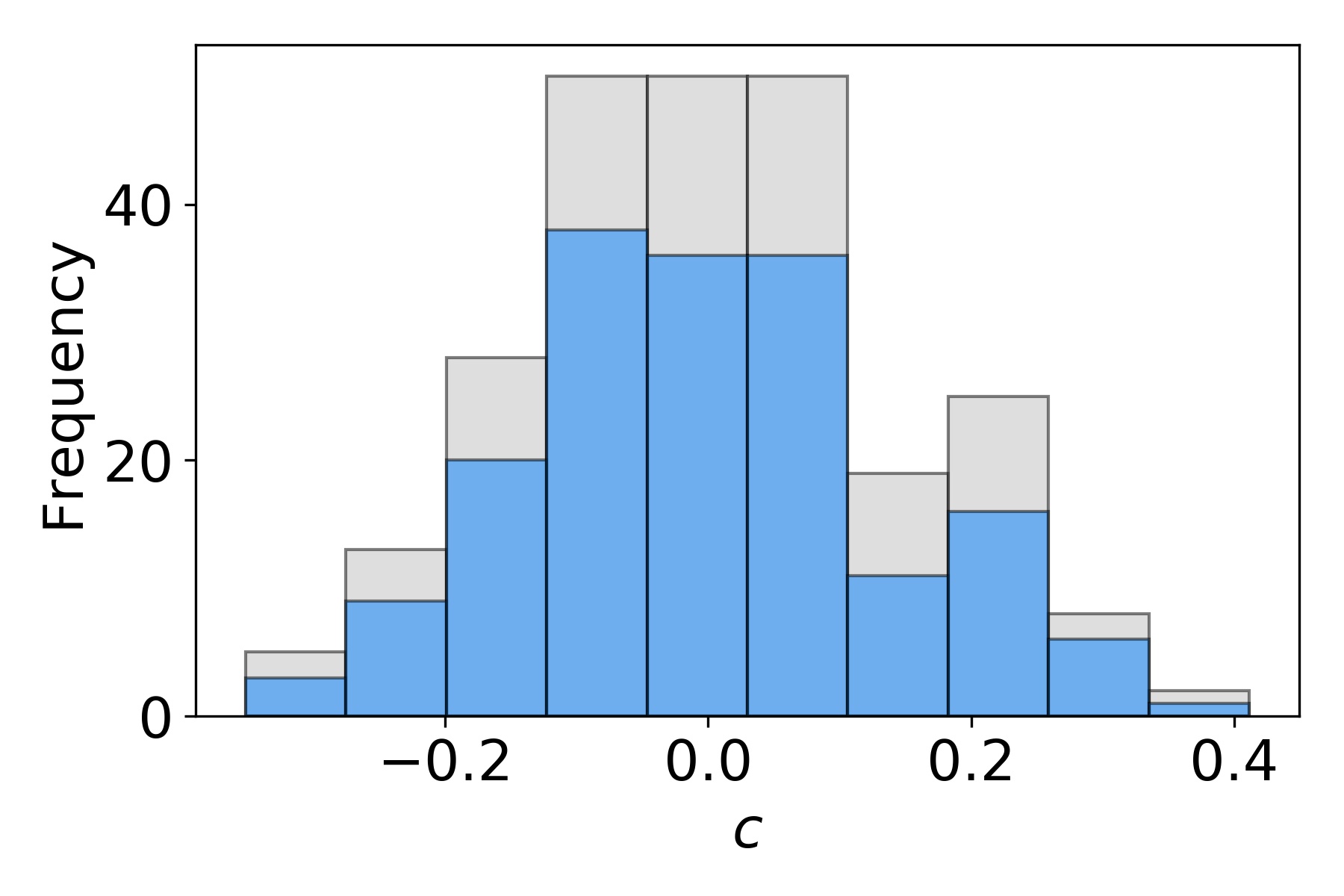} 
\caption{Histograms show the distributions of $m_b, x_1, c$, in a single simulated SALT-II data set. Grey histograms show the complete data set (observed and unobserved data points).  Blue histograms show truncated, observed data sets.  A truncation at a fixed point in $m_B$ causes data points to be truncated in $c,x_1$ in a way which is neither missing at random nor at a fixed truncation point.}
\label{fig:cosmo_data_scatter_2}
\end{figure*}

We will now demonstrate the validity of the SN cosmology posterior shown in Eq.~\ref{eq:SNBHMt1post} by testing the accuracy of its cosmological parameter inference using simulated supernova light curve fit parameters and redshifts.  We use the SN-BHM model is described in \cite{2011MNRAS.418.2308M} and \cite{2014MNRAS.437.3298M}, with fiducial parameters described in Table~\ref{tab:simsn}. 100 simulated data sets were created, and a cut was imposed on the peak $m_B$ magnitude such that data points with $m_B>24.0$ were discarded to produce truncated data sets, this cut-off was chosen to be loosely aligned with the $m_B$ truncation point in the Dark Energy Survey \citep{} 
For example plots of simulated SN data sets, see Figs.~\ref{fig:cosmo_data_scatter_malmquist}, \ref{fig:cosmo_data_scatter_1} and \ref{fig:cosmo_data_scatter_2}. As for the numerical trials with the Gaussian linear model, the posterior probability distribution for the truncated data set, given in Eq.~\ref{eq:SNBHMt1post} was sampled using the Multinest sampling algorithm and associated code \citep{2009MNRAS.398.1601F} with the pyMultinest wrapper \citep{2014A&A...564A.125B}. Resulting samples were processed using Chain Consumer \citep{Hinton2016}.

\subsection{Results}
\begin{figure*}
\centering
\includegraphics[width=1.0\linewidth]{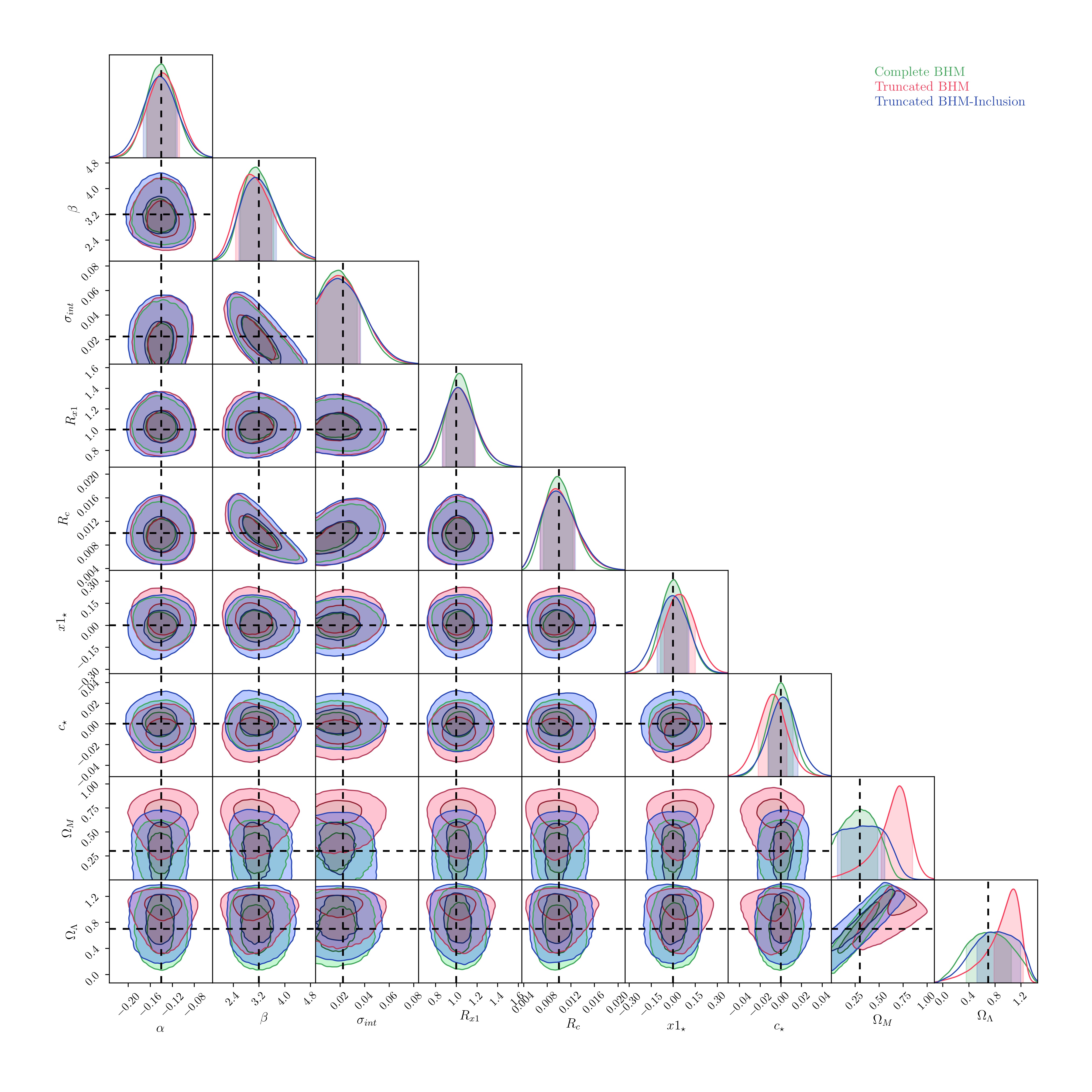}
\caption{Cosmological and supernova parameter inference from simulated SNIa data, fitted to a $\Lambda$CDM  model. Inner and outer contours enclose $68.3\%$ and $95\%$ of the posterior mass respectively. Green contours show complete (observed and unobserved) data set fitted with the standard Bayesian hierarchical model. Red contours show the truncated data sets fitted with the standard Bayesian hierarchical model, blue contours show the truncated data sets fitted with the extended Bayesian hierarchical model with inclusion model, that accounts for the truncation. This plot shows stacked results from 100 realizations of simulated data.
 }
\label{fig:lcdm_inference}
\end{figure*}

\begin{figure*}
\centering
\includegraphics[width=1.0\linewidth]{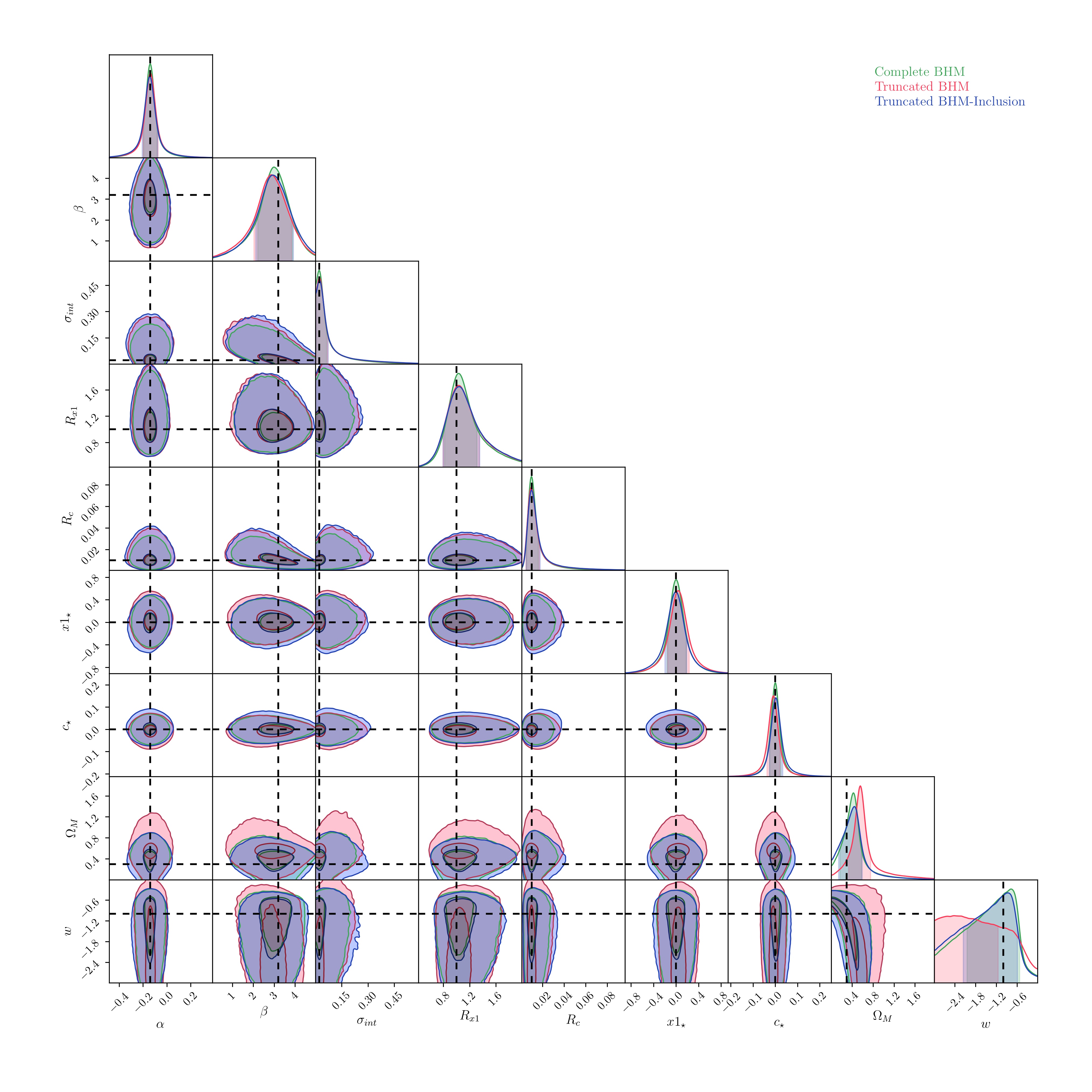}
\caption{Cosmological and supernova parameter inference from simulated SNIa data, fitted to a wCDM model. Inner and outer contours enclose $68.3\%$ and $95\%$ of the posterior mass respectively. Green curves show complete (observed and unobserved) data set fitted with the basic Bayesian hierarchical model. Red curves show the truncated data sets fitted with the basic Bayesian hierarchical model. Blue contours show the truncated data sets fitted with the extended Bayesian hierarchical model with inclusion model, that accounts for the truncation. This plot shows stacked results from 100 realizations of simulated data.}
\label{fig:wcdm_inference}
\end{figure*}

\begin{figure*}
\centering
\includegraphics[width=1.0\linewidth]{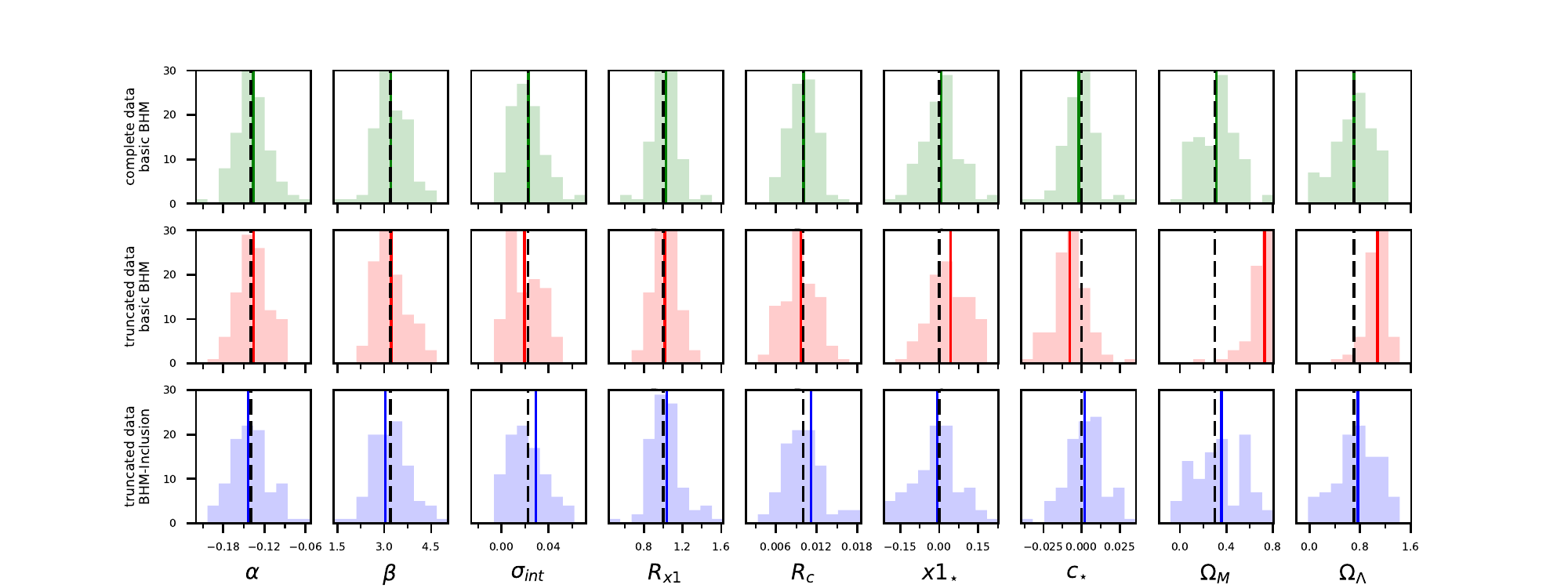} 
\includegraphics[width=1.0\linewidth]{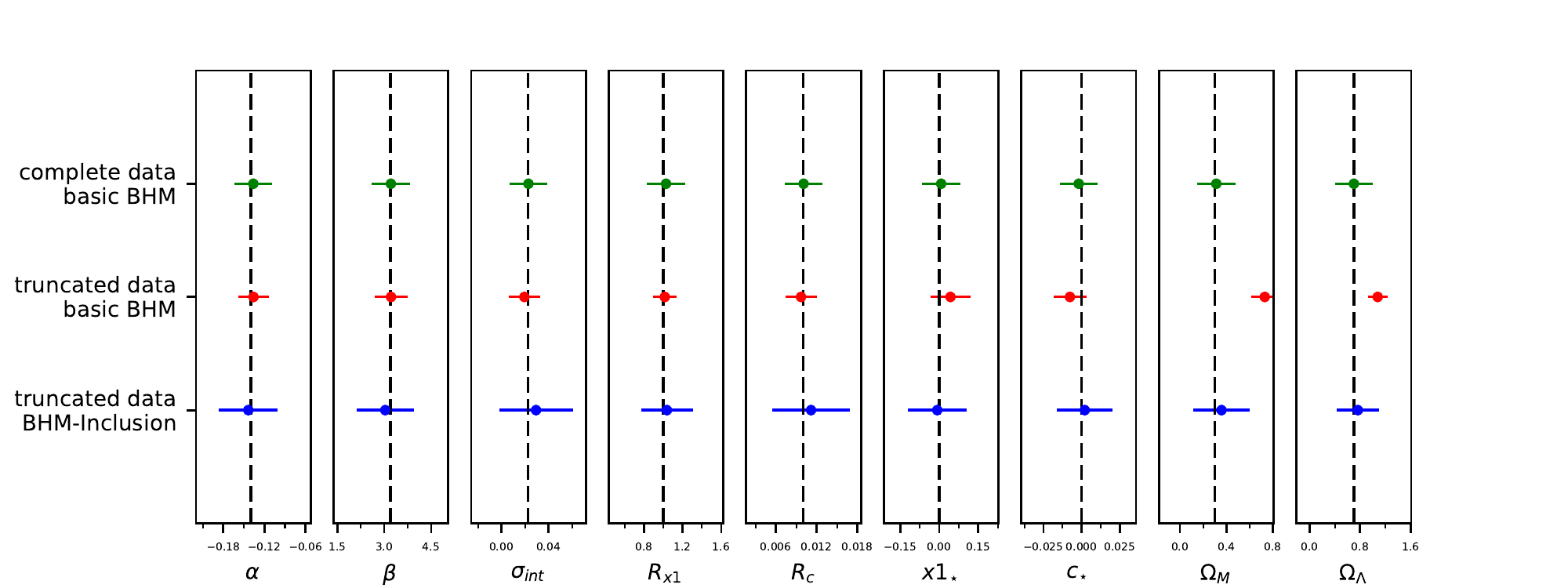} \vfill
\caption{Parameter inference in $\Lambda$CDM model.  In the upper plots, each histogram shows the point estimator (in this case the median) for each of 100 trials.  dashed black lines show the true values, solid coloured lines show the means of the point estimators.  Top row shows basic Bayesian hierarchical model applied to complete data set (which we can think of as the ideal situation that we wish to recover).  Middle row shows what happens when we attempt to use the basic method to recover the parameters of interest, showing the tendency towards a matter dominated universe.  Bottom row shows the results of using the Bayesian hierarchical model with inclusion model to infer parameters from the truncated data set, showing that the parameters are accurately recovered. The lower plots show the mean and standard deviations of the point estimators from the corresponding histograms, highlighting the accuracy and the precision with which the parameters are inferred in the truncated data set when using SNBHM with inclusion model.}
\label{fig:cosmo_hist_medians_lcdm}
\end{figure*}

\begin{figure*}
\centering
\includegraphics[width=1.0\linewidth]{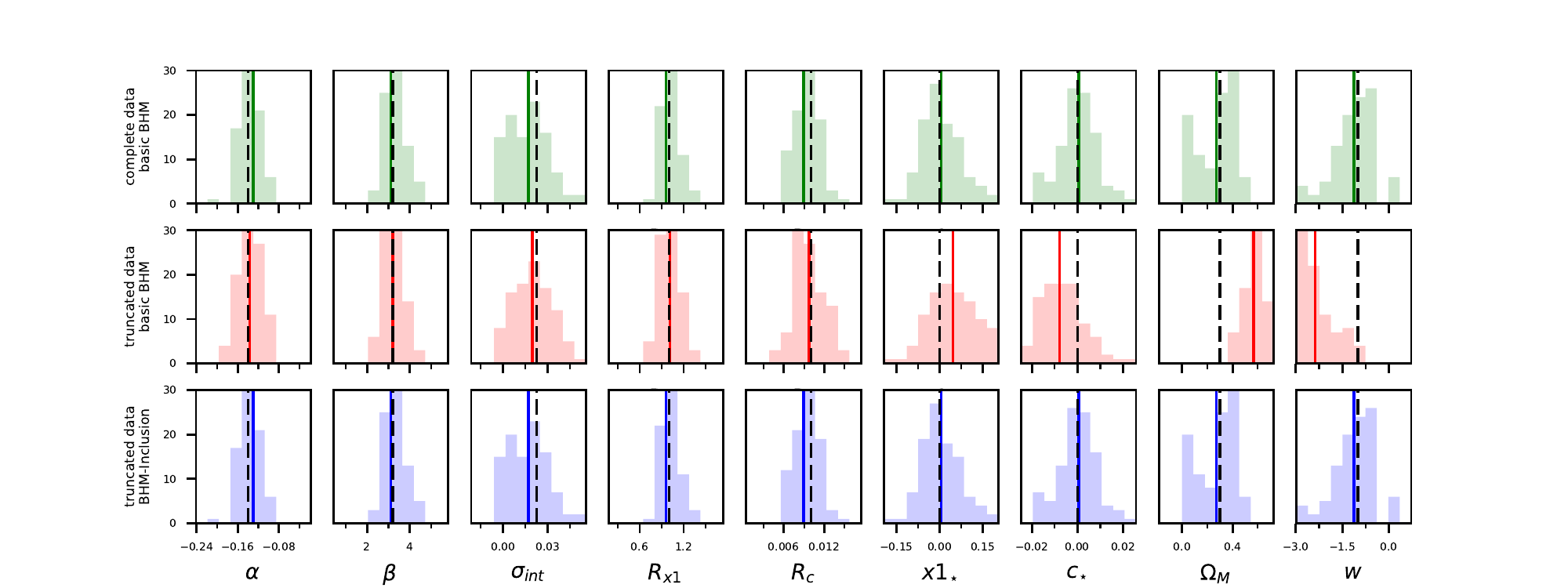} 
\includegraphics[width=1.0\linewidth]{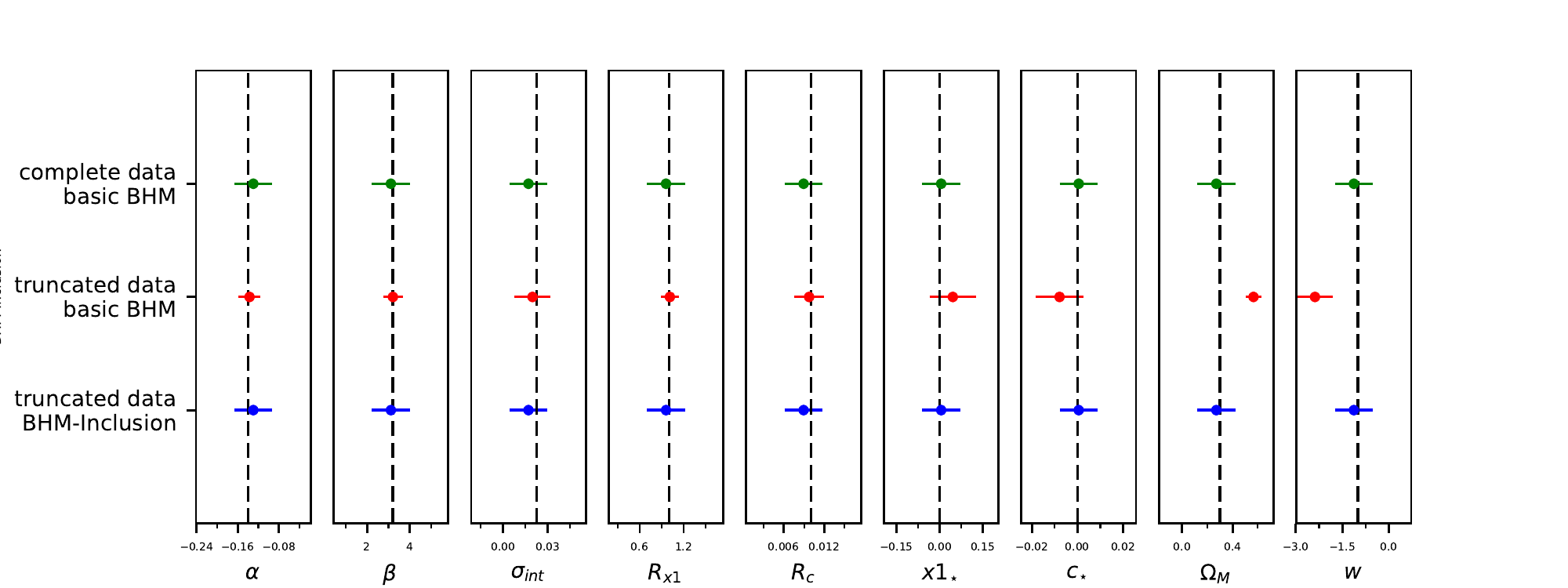} \vfill
\caption{Parameter inference in flat $w$CDM model.  In the upper plots, each histogram shows the point estimator (in this case the median) for each of 100 trials.  Dashed black lines show the true values, solid coloured lines show the mean of the point estimators.  Top row shows basic Bayesian hierarchical model applied to complete data set (which we can think of as the ideal situation that we wish to recover).  Middle row shows what happens when we attempt to use the basic method to recover the parameters of interest, showing the tendency towards a matter dominated universe.  Bottom row shows the results of using the Bayesian hierarchical model with inclusion model to infer parameters from the truncated data set, showing that the parameters are accurately recovered. The lower plots show the mean and standard deviations of the point estimators from the corresponding histograms, highlighting the accuracy and the precision with which the parameters are inferred in the truncated data set when using SNBHM with inclusion model.}
\label{fig:cosmo_hist_medians_wcdm}
\end{figure*}

The results shown in the stacked contour plots Fig.~\ref{fig:lcdm_inference} ($\Lambda$CDM), \ref{fig:wcdm_inference} (flat $w$CDM), and histogram plots Fig.~\ref{fig:cosmo_hist_medians_lcdm}, \ref{fig:cosmo_hist_medians_wcdm}, show that when the posterior for the basic SN-BHM Eq.~\ref{eq:GLM1post} is used to analyze the complete data set, all parameters are correctly inferred.  When the truncated data sets are analyzed using that method, we see expected biases in the recovery of the parameters of interest.  However, when the truncated data sets are analyzed using the Bayesian hierarchical model with the inclusion model, as described in Eq.~\ref{eq:GLM1postI} the parameters are again accurately inferred, with the expected slightly large uncertainties.  The larger interval estimators reflect the fact that there is increased uncertainty due to marginalizing over \emph{missing} data points.

\section{Discussion}
In this paper we have presented a derivation of the Bayesian posterior for truncated data sets, of the form of multidimensional Gaussian liner models, suitable for solving problems of parameter inference and model selection.  We have tested this methodology with a two dimensional Gaussian model.  We extended the general methodology to the specific case of SNIa cosmology analysis with the \salt2{} model and tested using basic simulations.  We have shown that the bias in parameter inference that would result from truncated data sets can be avoided by using a suitable inclusion model.  This paper presents the statistical derivation of the methodology and tests in in limited situations. The next steps will be to extend this methodology to include the intricacies of real data sets. We hope that this contribution allows others to develop their work in using  supernova data in problems of Bayesian model selection.


%
%

\twocolumn
\appendix 
\onecolumn

\section{Derivation of likelihood of Gaussian linear model with complete data set}
\label{sec:glmlike}
Following on from Eq.~\ref{eq:Bayes} in section \ref{sec:glmbayes}, the likelihood of the observed data can be expressed as
\begin{align}
\label{eq:like:glm}
p ( \mathbf{\xh}, \yh | \mathbf{a}, b, M ) &= \prod_i^N p(\mathbf{\xhi},\yhi |\mathbf{a},b,M) \nonumber \\
&= \iint d\mathbf{x}\, dy \,p(\mathbf{\xh},\yh | \mathbf{x}, y,\mathbf{a},b,M) \times p(y|\mathbf{x},\mathbf{a},b,M) \times p(\mathbf{x}|\mathbf{a},b,M) \, .
\end{align}
The probability of the observed data given the latent parameters is given by
\begin{align}
\label{eq:glmobsnoise}
p(\mathbf{\xh},\yh | \mathbf{x}, y,\mathbf{a},b,M) = \prod_i^N | 2\pi \Sigma_{C,i}^{-1}|^{-\frac{1}{2}} \exp{ ( -\frac{1}{2} ( 
(\mathbf{\whi} -\mathbf{w}_i)^T \Sigma_{c,i}      
(
\mathbf{\whi} -\mathbf{w}_i )
)} \, ,
\end{align}
where
\begin{align}
\mathbf{w}_i &= \begin{bmatrix} 
y_i \\ 
\bf{x}_i  \end{bmatrix} \in \mathbb{R}^{(1+J)} \, , \\
\mathbf{\whi} &= \begin{bmatrix} 
\yhi  \\ 
\bf{\xhi}   \end{bmatrix} \in \mathbb{R}^{(1+J)} \, , \\ 
\Sigma_{c,i} &=
\begin{bmatrix} 
\syi^2 & \mathbf{\syxi}\\
\mathbf{\syxi} & \mathbf{\sx}i^2 
\end{bmatrix}  \in \mathbb{R}^{(1+J) \times (1+J)} \, ,
\end{align}
and the probability of latent $y$ given the latent $\mathbf{x}$, slope $\mathbf{a}$ and intercept $b$ is
\begin{align}
\label{eq:glmintdisp}
p(y|\mathbf{x},\mathbf{a},b,M) =  \prod_i^N |2\pi \sinti^2|^{-\frac{1}{2}} \exp{ \left( -\frac{1}{2} \left( 
\frac{(y_i -(\mathbf{ax}_i +b))^2}{\sinti^2}  \right)  \right)} \,.
\end{align}
The expression $p(\mathbf{x}|a,b,M)$ describes the distribution of the latent parameters $\mathbf{x}$, which we shall represent as a Gaussian (for a discussion on why a Gaussian is the preferred choice, see e.g., \cite{Gull1989}, \cite{2011MNRAS.418.2308M}, and \cite{D'Agostini:1995fv}). 
%

\begin{align}
&\normal {\mathbf{x}_ i} {\mathbf{\xnot}} {\mathbf{\Rx}} \, , \\
&\mathbf{\xnot} \in \mathbb{R}^J \, , \nonumber \\
&\mathbf{\Rx} \in \mathbb{R}^{J \times J} \, , \nonumber
\end{align}
such that we can write the prior probability of the latent $\mathbf{x}$ variables as follows, marginalizing over the mean and standard deviation of the parent distribution:
\begin{align}
\label{eq:glmlatentprob}
p(\mathbf{x}|,\mathbf{a},b,M) &= \iint d\mathbf{\Rx} \, d\mathbf{\xnot} \,p(\mathbf{x}|\mathbf{\Rx},\mathbf{\xnot},\mathbf{a},b,M) \nonumber \\
&=  \iint d\mathbf{\Rx} \, d\mathbf{\xnot}  \prod_i^N |2\pi \mathbf{\Rx}^{2}|^{-\frac{1}{2}} \exp{ \left( -\frac{1}{2}  (\mathbf{x}_i -\mathbf{\xnot})^T\mathbf{\Rx}^{-2} (\mathbf{x}_i -\mathbf{\xnot})     
\right)} \, .
\end{align}
Re-writing Eqs.~\ref{eq:glmobsnoise},~\ref{eq:glmintdisp},~\ref{eq:glmlatentprob}, the three Gaussian terms in Eq.~\ref{eq:like:glm} as a single Gaussian in $\mathbf{x},y$, and integrating over the latent $\mathbf{x},{y}$ gives:
\begin{align}
\label{eq:2dbhm}
p ( \mathbf{\xh}, \yh | \mathbf{a}, b, M ) &=  \iint  d\mathbf{\Rx} \, d\mathbf{\xnot} \, \prod_i^N   | 2\pi \Sigma_{v,i}|^{-\frac{1}{2}} \exp{ \left( -\frac{1}{2} \left( 
\left(\mathbf{\whi} -\mathbf{q}_i\right)^T \Sigma_{v,i}^{-1}      
\left(
\mathbf{\whi} -\mathbf{q}_i \right)
\right)\right)} \, ,
\end{align}
where
\begin{align}
\mathbf{q}_i &= \begin{bmatrix} 
b + \mathbf{a^T \xnot}  \\ 
\mathbf{\xnot}   \end{bmatrix}  \in \mathbb{R}^{(1+J)} \,, \\
\Sigma_{v,i} &=
\begin{bmatrix} 
\syi^2 + \sint^2 + \mathbf{a}^T\mathbf{\Rx}^2\mathbf{a} & \mathbf{\syxi}^T  +\mathbf{a}^T\mathbf{\Rx}^2  \\
\syxi  + \mathbf{\Rx}^2 \mathbf{a} & \mathbf{\sxi}^2 +\mathbf{\Rx}^2 
\end{bmatrix}  \in \mathbb{R}^{(1+J) \times (1 +J)} \,,
\end{align}
Substituting Eq.~\ref{eq:2dbhm} into the Bayes equation, Eq.~\ref{eq:Bayes}, gives us the expression for the posterior probability of the parameters of interest (in this case the slope and intercept parameters, $\mathbf{a}$ and $b$), given the observed data, $\mathbf{\xh}$, $\yh$. For questions of parameter inference we can drop the normalizing constant given by the Bayesian evidence to give the posterior
\begin{align}
p ( \mathbf{a},b |\mathbf{\xh}, \yh,I,M )
&\propto \iint   \, d\mathbf{\Rx} \, d\mathbf{\xnot} \, \prod_i^{N}  \left| 2\pi \Sigma_{v,i}\right|^{-\frac{1}{2}} \exp{ \left( -\frac{1}{2} \left( (\mathbf{\whi} -\mathbf{q}_i)^T \Sigma^{-1}_{v,i} (\mathbf{\whi} -\mathbf{q}_i )\right)\right)} \nonumber \\
&\times  p(\mathbf{\xnot,\Rx}|I,M)  p(\mathbf{a},b|I,M)  \label{eq:GLM1post} \, .
\end{align}
%

%
\subsection{Derivation of likelihood of Gaussian linear model with truncated data set}
\label{sec:glmtrunc}
%
To find the probability of the observed data points, we need to marginalize over the missing data points
\begin{align}
p ( \mathbf{\xh}^{\text{obs}}, \yh^{\text{obs}} | \mathbf{a}, b,y^{\text{thresh}}, I, M ) &= \iint d\mathbf{\xh}^{\text{mis}} d\yh^{\text{mis}} p(\mathbf{\xh}^{\text{obs}},\yh^{\text{obs}},\mathbf{\xh}^{\text{mis}},\yh^{\text{mis}} | \mathbf{a},b,I,M)  \\
&= \prod_i^{N-m} p(\mathbf{\xhi}^{\text{obs}},\yhi^{\text{obs}} |\mathbf{a},b,I=1,M) \prod_i^m \int_{-\infty}^{+\infty}\int_{y^{\text{thresh}}}^{+\infty} d\mathbf{\xhi}^{\text{mis}} d\yhi^{\text{mis}} p(\mathbf{\xhi}^{\text{mis}},\yhi^{\text{mis}} |\mathbf{a},b,I=0,M) \, .
\label{eq:like:trunc1}
\end{align}
The two probability terms in Eq.~\ref{eq:like:trunc1} have the same form as  Eq.~\ref{eq:2dbhm}, the only difference being that the first  term is \emph{evaluated} over the \emph{observed} data points, and the second term is \emph{marginalized} over the \emph{missing} data points.  The Binomial factor accounts for the different possible combinations of total and observed data points.  For the censored case where $N$ is known
\begin{align}
p ( \mathbf{\xh}^{\text{obs}}, \yh^{\text{obs}} | \mathbf{a}, b,I,N,M )
&=  \iint  d\mathbf{\Rx} \, d\mathbf{\xnot} \,  \binom{N}{\Nobs} \prod_i^{\Nobs}   | 2\pi \Sigma^{\text{obs}} _{v,i}|^{-\frac{1}{2}} \exp{ \left( -\frac{1}{2} \left( (\mathbf{\whi}^{\text{obs}} -\mathbf{q}_i)^T \Sigma^{\text{obs}^{-1}} _{v,i} (\mathbf{\whi}^{\text{obs}} -\mathbf{q}_i )\right)\right)} \nonumber \\
& \times \prod_i^{m} \int_{-\infty}^{+\infty}\int_{y^{\text{thresh}}}^{+\infty} d\mathbf{\xhi}^{\text{mis}} d\yhi^{\text{mis}}  | 2\pi \Sigma^{\text{mis}} _{v,i}|^{-\frac{1}{2}} \exp{ \left( -\frac{1}{2} \left( (\mathbf{\whi}^{\text{mis}} -\mathbf{q}_i)^T \Sigma^{\text{mis}^{-1}} _{v,i} (\mathbf{\whi}^{\text{mis}} -\mathbf{q}_i )\right)\right)} \, .
\label{eq:like:trunc2}
\end{align}
For the truncated case where $N$ is unknown we use a Jeffreys' prior on $N$ 
\be
p(N | \mathbf{a},b,I,M)  = \frac{1}{N} \, ,
\ee
and marginalize over all possible values of $N$, such that
\begin{align}
p ( \mathbf{\xh}^{\text{obs}}, \yh^{\text{obs}} | \mathbf{a}, b,I,M )
&= \int_{\Nobs}^{\infty} dN \, p ( \mathbf{\xh}^{\text{obs}}, \yh^{\text{obs}} | \mathbf{a}, b,I,N,M ) p(N | \mathbf{a},b,I,M) \\
&= \int_{\Nobs}^{\infty} dN \, \iint   \, d\mathbf{\Rx} \, d\mathbf{\xnot} \, \frac{1}{N} \binom{N}{\Nobs} \prod_i^{\Nobs}   | 2\pi \Sigma^{\text{obs}} _{v,i}|^{-\frac{1}{2}} \exp{ \left( -\frac{1}{2} \left( (\mathbf{\whi}^{\text{obs}} -\mathbf{q}_i)^T \Sigma^{\text{obs}^{-1}} _{v,i} (\mathbf{\whi}^{\text{obs}} -\mathbf{q}_i )\right)\right)} \nonumber \\
& \times \prod_i^{m} \int_{-\infty}^{+\infty}\int_{y^{\text{thresh}}}^{+\infty} d\mathbf{\xhi}^{\text{mis}} d\yhi^{\text{mis}}  | 2\pi \Sigma^{\text{mis}} _{v,i}|^{-\frac{1}{2}} \exp{ \left( -\frac{1}{2} \left( (\mathbf{\whi}^{\text{mis}} -\mathbf{q}_i)^T \Sigma^{\text{mis}^{-1}} _{v,i} (\mathbf{\whi}^{\text{mis}} -\mathbf{q}_i )\right) \right)} \label{eq:GLM1like} \, .
\end{align}
Substituting Eq.~\ref{eq:GLM1like} into the Bayes equation, Eq.~\ref{eq:Bayes}, gives us the expression for the posterior probability of the parameters of interest (in this case the slope and intercept parameters, $\mathbf{a}$ and $b$), given the observed data, $\mathbf{\xh}$, $\yh$ for a truncated data set where an unknown number of observations are missing.  For questions of parameter inference we can drop the normalizing constant given by the Bayesian evidence to give the posterior:
\begin{align}
p ( \mathbf{a},b |\mathbf{\xh}^{\text{obs}}, \yh^{\text{obs}}, y^{\text{thresh}}I,M )
&\propto \int_{\Nobs}^{\infty} dN \, \iint   \, d\mathbf{\Rx} \, d\mathbf{\xnot} \, \frac{1}{N} \binom{N}{\Nobs} \prod_i^{\Nobs}   | 2\pi \Sigma^{\text{obs}} _{v,i}|^{-\frac{1}{2}} \exp{ \left( -\frac{1}{2} \left( (\mathbf{\whi}^{\text{obs}} -\mathbf{q}_i)^T \Sigma^{\text{obs}^{-1}} _{v,i} (\mathbf{\whi}^{\text{obs}} -\mathbf{q}_i )\right)\right)} \nonumber \\
& \times \prod_i^{m} \int_{-\infty}^{+\infty}\int_{y^{\text{thresh}}}^{+\infty} d\mathbf{\xhi}^{\text{mis}} d\yhi^{\text{mis}}  | 2\pi \Sigma^{\text{mis}} _{v,i}|^{-\frac{1}{2}} \exp{ \left( -\frac{1}{2} \left( (\mathbf{\whi}^{\text{mis}} -\mathbf{q}_i)^T \Sigma^{\text{mis}^{-1}} _{v,i} (\mathbf{\whi}^{\text{mis}} -\mathbf{q}_i )\right)\right)} \nonumber \\ 
&\times p(\mathbf{a},b|I,M)  p(\mathbf{\xnot,\Rx}|I,M)  \, .
\label{eq:GLM1postI}
\end{align}
%

\section{Acknowledgments}
MCM and MS were supported by DOE grant DE-FOA-0001358 and NSF grant AST-1517742. We thank Roberto Trotta and Rick Kessler for helpful discussions regarding this work.
\label{lastpage}
\bibliographystyle{mnras}
\bibliography{BasicBiblio}
\end{document}